\documentclass[a4paper]{article}
\usepackage{a4wide}
\usepackage{graphicx}
\usepackage[dvipsnames]{xcolor}
\usepackage{caption}
\usepackage{empheq}
\usepackage{wrapfig}
\usepackage{amsthm}
\usepackage{amssymb, latexsym, mathrsfs}
\usepackage{amsmath}
\usepackage{dsfont}
\usepackage{enumerate}
\usepackage{algorithm}
\usepackage{color,soul}
\usepackage{transparent}
\usepackage{subfig}
\usepackage{url}
\usepackage[affil-it]{authblk}
\usepackage{algorithm}
\usepackage{algpseudocode}
\usepackage{booktabs}
\usepackage{multirow}
\usepackage{float} 
\usepackage[showframe=false]{geometry}
\usepackage{enumitem}
\usepackage{graphicx}
\usepackage{epsfig}
\usepackage{pgfplots}
\newlength\fheight
\newlength\fwidth
\usepackage{indentfirst}
\pgfplotsset{compat=newest}
\pgfplotsset{plot coordinates/math parser=false}
\usepackage{amsmath,amsfonts,amssymb,amsthm}

\newtheorem{rmk}{Remark}[section]

\date{}

\usepackage{amsthm}
\usepackage{amssymb}
\theoremstyle{plain}

\theoremstyle{definition}

\begin{document}
	\title{\LARGE \bf Supervised MPC control of large-scale electricity networks via clustering methods}
	
	\author[1]{Alessio La Bella}
	\author[2]{Pascal Klaus}
	\author[2]{Giancarlo Ferrari-Trecate}
	\author[1]{Riccardo Scattolini}
	
	\affil[1]{\small Dipartimento di Elettronica, Informazione e Bioingegneria, Politecnico di Milano, Milano,  Italy \\ (e-mail: alessio.labella@polimi.it, riccardo.scattolini@polimi.it)}
	\affil[2]{\small Automatic Control Laboratory, \'Ecole Polytechnique F\'ed\'erale de Lausanne, Lausanne, Switzerland \\(e-mail: pascal.klaus@alumni.epfl.ch,\; giancarlo.ferraritrecate@epfl.ch)}
     \date{\textbf{Technical Report}\\ April, 2020}
\maketitle
     \begin{abstract}{This paper describes a control approach for large-scale electricity networks, with the goal of efficiently coordinating distributed generators to balance unexpected load variations with respect to nominal forecasts. To mitigate the difficulties due to the size of the problem, the proposed methodology is divided in two steps. First, the network is partitioned into clusters, composed of serveral dispatchable and non dispatchable generators, storage systems, and loads. A clustering algorithm is designed with the aim of obtaining clusters with the following characteristics: (\emph{i}) they must be compact, keeping the distance between generators and loads as small as possible; (\emph{ii}) they must be able to internally balance load variations to the maximum possible extent. Once the network clustering has been completed, a two layer control system is designed. At the lower layer, a local Model Predictive Controller is associated to each cluster for managing the available generation and storage elements to compensate local load variations. If the local sources are not sufficient to balance the cluster's load variations, a power request is sent to the supervisory layer, which optimally distributes additional resources available from the other clusters of the network. To enhance the scalability of the approach, the supervisor is implemented relying on a fully distributed optimization algorithm. The IEEE 118-bus system is used to test the proposed design procedure in a non trivial scenario.}
     \end{abstract}

\section{Introduction}
\begin{figure}[t!]
	\centering
	\includegraphics[width=0.7\linewidth]{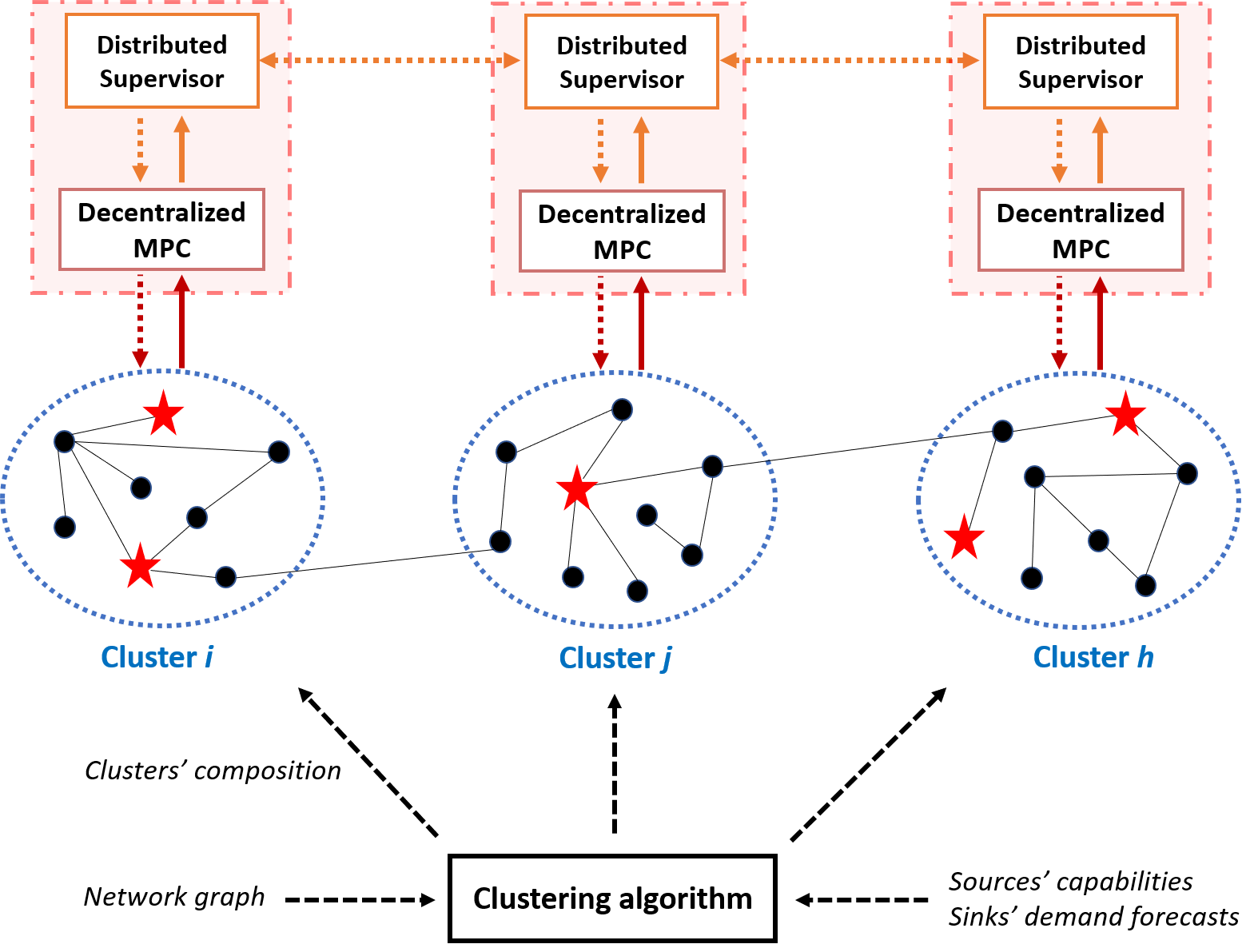}
	\caption{Overall control architecture applied to a partitioned network with sources (red stars) and sinks (black dots).}
	\label{fig:controlscheme}
\end{figure}

With the increasing penetration of volatile renewable energy sources in the electrical grid, such as photovoltaic  and wind-turbine generators, and of new non-deterministic loads, such as charging stations for electric vehicles, grid control has become an increasingly complex task. Indeed, the output power of renewables and the load requests frequently deviate from the nominal forecasts, causing continuous power imbalances between generation and absorption which must be promptly restored in order to avoid undesired and critical deviations of the network frequency \cite{hirth2015balancing}.
The control strategies established in the past heavily rely on the inertia of large rotating generators to ensure stability and on-demand power generation. As such, they are inadequate for coping with the above challenges. In recent years, the development of new solutions for grid control has received increasing attention and many different approaches have been proposed with the aim of ensuring the power balance by compensating unexpected power variations of renewable energy sources and loads. A control strategy to coordinate distributed generators to counteract power variations of external loads is proposed in \cite{hong2017controlling}. In \cite{majzoobi2017application}, a micro-grid equipped with several distributed generators is controlled to compensate the power variability of multiple power consumers, all connected to the same distribution feeder. The availability of storage systems, together with reliable forecasts on the generation from renewable sources
and on the consumption of loads, makes Model Predictive Control (MPC) the most promising control design method for this class of problems \cite{maciejowski2002predictive}.  The use of centralized MPC regulators for balancing unexpected power variations is discussed in \cite{hug2010coordination,cominesi2017two}. However, pure centralized approaches suffer from scalability and computational complexity issues, and therefore they are not advisable to efficiently control large-scale grids.\\
An effective solution to control large-scale networks consists in first splitting them in small-scale non-overlapping sub-networks, and then optimizing their internal operations and power exchanges. Different methods are available in the literature for network partitioning, mainly based on topological properties of the associated network graph \cite{4410510,cortes2018microgrid,cotilla2013multi}. Nevertheless, to efficiently partition a network, not only its topology, but also the effective power capability of distributed generators and the load power profiles should be taken into account. A simple online strategy to create partitions with enough power capability to balance unforeseen load variations in isolated DC micro-grids is designed in \cite{martinelli2019secondary}. \\
Concerning control architectures for large-scale distribution networks divided into sub-areas, distributed optimization-based control schemes are proposed in \cite{hug2009decentralized},  \cite{baker2016distributed}. However, these methods are based on slow and iterative procedures, which are not advisable to quickly compensate power imbalances and to coordinate the power exchanges between the different grid areas.
\\
Given the mentioned issues, the contribution of this present work consists in the design of a multi-layer control architecture according to a \emph{divide et impera} strategy ensuring efficient network partitioning and prompt compensation of power imbalances.
Precisely, the proposed control design strategy relies on the following steps:

\begin{itemize}
	\item The electricity network is first partitioned into small-scale clusters of nodes, composed of distributed generators, either dispatchable or non-dispatchable, and loads. These units are referred in the following as sources and sinks, respectively. The design of the clusters has a twofold objective. First, they must be "compact", keeping the distance between any individual pair of nodes within each cluster as small as possible. Second, they must have the highest possible degree of independence, i.e. they must be able of internally balancing load power variations using the available generators, requiring for external assistance just in a few unlikely and extreme scenarios.
	\item The partitioned network is regulated by a two layer control scheme. The lower layer is composed by decentralized Model Predictive Control (MPC) regulators, implemented at the top of each grid cluster,
	and coordinating dispatchable generators and associated storage units in each cluster to balance the local demand variability. If local sources are not sufficient to balance a cluster, the corresponding local MPC regulator can issue a power request to the upper supervisory layer, which is in charge of supporting those clusters that are subject to shortages by redirecting resources from the others. The supervisory layer is designed through a fully distributed optimization problem, based on the distributed Dual Consensus ADMM (DC-ADMM) algorithm, see \cite{chang2016proximal}. This two layer scheme generalizes the preliminary results reported in \cite{bib:LaBella2019}. A schematic of the proposed two-layer control architecture is depicted in Figure \ref{fig:controlscheme}.
\end{itemize}

\section{Network clustering algorithm} \label{sec:clustering}
\begin{figure}[b!]
	\centering
	\includegraphics[width=0.7\linewidth]{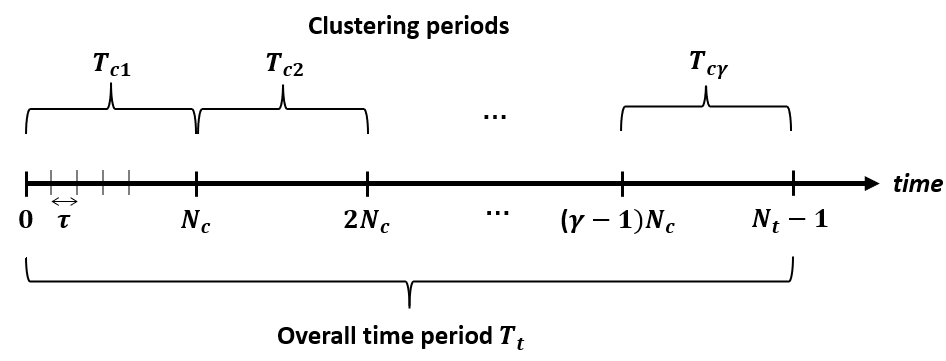}
	\caption{Schematic of time periods in which clustering and control algorithms take action. }
	\label{fig:timeperiods}
\end{figure}

We denote by $\tau$ the basic sampling period used in the clustering and control algorithms. Clusters and control actions are defined for an overall time period composed of $N_t$ sampling time instants and denoted as $T_t=\{0,\hdots,N_t-1\}$.
To exploit recent information about load absorption and renewable production, every $N_c$ time steps, the network is periodically re-partitioned into clusters. For simplicity, $N_c$ is defined as a divisor of $N_t$, and the term $\gamma = N_t/N_c$ denotes the number of \emph{clustering periods} over $T_t$. Considering $\eta =1,\hdots, \gamma$, the $\eta$-th clustering period is composed by the time instants \mbox{$T_{c \eta} =\{(\eta-1)N_c,\hdots,\eta N_c-1\}$}. A representation of the time division is depicted in Figure \ref{fig:timeperiods}.

\pagebreak
To partition the network into clusters, \emph{for each clustering period} the following procedure is performed.
\begin {itemize}
\item The first step consists in associating sources to sinks (i.e. generators to loads), defining the optimal power transactions to compensate power imbalances caused by unexpected load variations. This is pursued through suitably defined optimization problems, considering also constraints related to the capability of sources and the demand trends of sinks.
Since the network is assumed to be connected, each source can be associated to any sink of the network. Moreover, each source can be associated to more than one sink, and each sink demand can be satisfied by multiple source nodes.
\item Once the transactions between sources and sinks have been identified, these are projected onto the real network graph. As it will be explained in the next paragraphs, the transactions are mapped considering the shortest path connecting a source and a sink, in order to maximize the compactness of the clusters. The projection of transactions into the real network serves to define the importance of each transmission line by properly computing suitable weights.
\item Finally, the network graph is partitioned by minimising the edge-cut, i.e. the sum of the weights of the edges connecting individual clusters. This is a specific instance of the so-called \textit{k-way} partitioning problem, in which a given graph is divided into a pre-determined amount of balanced, connected and non-overlapping clusters \cite{karypis2000multilevel}. This task is performed using the well-known software tool METIS \cite{bib:METIS_doc}, designed for graph partitioning problems.
\end{itemize}
The electric network is modeled as a \textit{undirected}, \textit{connected} and \textit{weighted} graph $\mathcal{G}(\mathcal{V},\mathcal{E})$, where $\mathcal{V}=\{1,\dots,V\}$ indicates the set of nodes, while $\mathcal{E} \subseteq \mathcal{V} \times \mathcal{V}$ is the set of edges. We denote by $|\mathcal{V}|=V$ the cardinality of the set of nodes. Each edge $e\in \mathcal{E}$ is characterized by a weight $l_{e}$, for instance representing the physical length of the corresponding network line. For the sake of clarity, each edge $e\in \mathcal{E}$ can be also represented by the pair of its end-nodes $a,b \in \mathcal{V}$, i.e. $[a,b]=e $.
Since the graph is assumed to be connected, $ \forall \,i,\,j\in \mathcal{V}$, there exists at least one path  connecting the nodes $i$ and $j$. The cost of a path $i_0,i_1,\hdots,i_n$ from node $i_0$ to node $i_n$ is $\sum_{k=1}^{n}l_{[i_{k-1},i_{k}]} $. A path from $i_0$ to $i_n$ is termed the \textit{shortest path} if the associated cost is minimal among all paths from $i_0$ to $i_n$.
The set of edges composing the shortest path connecting two different nodes $i,\,j \in \mathcal{V}$ is denoted by $\mathcal{L}_{ij} \subseteq \mathcal{E} $ and the associated cost is
\begin{equation}\label{eq:cost_transaction}
c_{ij} \;=\! \sum_{e\, \in\, \mathcal{L}_{ij} } \hspace{0mm}l_{e}\;.
\end{equation}

It is worth noticing that there may be multiple shortest paths connecting $i$ and $j$; in this case, $\mathcal{L}_{ij}$ is chosen arbitrarily among these paths.
Given two nodes in a graph, several algorithms for computing the shortest paths are available in the literature, such as the Dijkstra's algorithm \cite{dijkstra1959note}. Therefore, given the network graph $\mathcal{G}(\mathcal{V},\mathcal{E})$, it is assumed that for each pair of nodes $i,j\in\mathcal{V}$, the shortest path $\mathcal{L}_{ij}$ is already available, together with the cost terms $c_{ij}$.
\\
\\
As mentioned, the network graph comprises source nodes and sink nodes, which are included in the sets $\mathcal{S} \subseteq \mathcal{V}$ and  $\mathcal{D} \subseteq \mathcal{V}$, respectively. Moreover, consistently with many realistic applications, it is supposed that some source nodes are also endowed with storage devices, e.g. batteries, which are included in the set $\mathcal{B} \subseteq \mathcal{S}$.  For the sake of simplicity, storages are not considered during the clustering procedure, as the evolution of the states of charges cannot be known a-priori; instead, their presence will be exploited during the control design phase described in Section \ref{sec:two_layer_control}.
\\ \\
Without loss of generality, consider now the first clustering period, i.e. the one covering the time interval $T_c=\{0,\hdots,N_c-1\}$, and the time index $k$ denoting the generic sampling time in ${T}_c$. At any $k$, and for each source $i\in\mathcal{S}$, denote by ${S}_i^{\uparrow}(k)$ and ${S}_i^{\downarrow}(k)$ its maximum and minimum power limits, respectively. Analogously, for each sink  $j\in \mathcal{D}$, the maximum and minimum power absorption are ${D}_i^{\uparrow}(k)$ and ${D}_i^{\downarrow}(k)$. In order to consider the variability of sources and sinks within the clustering period, these limits are assumed to depend upon time.
\\
In addition, we assume that for any time instant $k$ an Optimal Power Flow (OPF) problem has been solved in nominal conditions, so that given the nominal values of loads' absorption, i.e. $\bar{D}_j(k)\in [D_i^{\downarrow}(k),D_i^{\uparrow}(k) ]$, the corresponding optimal nominal values $\bar{S}_i(k)\in [S_i^{\downarrow}(k),S_i^{\uparrow}(k) ]$ are computed.
Therefore, at any $k$ the maximum and minimum permissible power variations of generators and loads with respect to their nominal values can be defined as\\
\begin{alignat}{4}
\bar{s}_i^{\uparrow}(k)    \;&=&&  \;S_i^{\uparrow}(k)   - \bar{S}_i(k)&&&\geq 0
\label{eq:up_gen}\\
\bar{s}_i^{\downarrow}(k)  \;&=&&  \;S_i^{\downarrow}(k) - \bar{S}_i(k)&&&\leq 0 \label{eq:down_gen}\\
\bar{d}_j^{\uparrow}(k)    \;&=&&  \;D_j^{\uparrow}(k)   - \bar{D}_j(k)\;&&&\geq 0 \label{eq:up_demand}\\
\bar{d}_j^{\downarrow}(k)  \;&=&&  \;D_j^{\downarrow}(k) - \bar{D}_j(k)\;&&&\leq 0 \label{eq:down_demand}
\end{alignat}
\\
\subsection{Transactions between sources and sinks}\label{subsec:trans}
The first step of the procedure previously described consists in defining the power transactions between sources and sinks required to compensate power imbalances.
To this end, the variable $x_{ij}(k)$ is introduced to denote the power variation, with respect to its nominal value, flowing from source $i \in \mathcal{S}$ to sink $j \in\mathcal{D}$  at time $k$. Moreover, a slack variable $x^s_{j}$ is used to identify the amount of demand variation of sink $j$ that cannot be provided by sources in $ \mathcal{S}$. If the network to be partitioned is not isolated, e.g. it is a distribution electrical network connected to the main utility, $\sum_{j\in \mathcal{D}}x^s_{j}(k)$ corresponds to the total amount of demand not satisfied by internal sources, but by external entities.\\ \\
The transactions of flows between sources and sinks, i.e. the values of $x_{ij}$ can be computed according to different approaches. Here, for each pair $(i,j)$, we compute $x_{ij}$ as the "average" absolute value of the flows obtained by considering the optimal solutions of two different optimization problems, namely the ones corresponding to the cases where all the load power variations take either their maximum values ${d}_j^{\uparrow}(k)$ or their minimum ones ${d}_j^{\downarrow}(k)$. These power flows are named  $x^\uparrow_{ij}(k)$ and $x^\downarrow_{ij}(k)$, respectively. Accordingly, $x^{s \uparrow}_j(k)$ and $x^{s \downarrow}_j(k)$ will denote the amount of power variability of sink $j$ that cannot be satisfied by internal sources in the two considered scenarios.\\ \\
In view of the previous considerations, the following problems $P1$ and $P2$ are solved at each $k=0,\dots,N_c-1$ to compute the optimal values of $x^\uparrow_{ij}(k)$ and $x^\downarrow_{ij}(k)$, respectively, in the two scenarios.\\
\textbf{\emph{Problem P1}}
\begin{subequations}\label{eq:robust_up}
\begin{align}
\underset{\substack{x^\uparrow_{ij}( k),\,x^{s\uparrow}_{j}(k)\\}}{\text{min}}\;\; & \sum_{i \in \mathcal{S}} \sum_{j \in \mathcal{D}} c_{ij}  \vert\, x^\uparrow _{ij}(k) \,\vert +  \sum_{j \in \mathcal{D}} c_{s}\, \vert\,x^{s\uparrow}_{j}(k)\,\vert &&\\
\text{subject to} \\
&  \sum_{j \in \mathcal{D}}x^\uparrow_{ij}(k) \leq \bar{s}_i^\uparrow(k)\,,  &&\forall i \in \mathcal{S},\\
&  \sum_{j \in \mathcal{D}}x^\uparrow_{ij}(k) \geq \bar{s}_i^\downarrow(k)\,,  &&\forall i \in \mathcal{S},\\
& \sum_{i \in \mathcal{S}}x^\uparrow_{ij}(k) + x^{s\uparrow}_{j}(k) = \bar{d}^{\,\uparrow}_j(k)\,,   &&\forall j \in \mathcal{D},
\end{align}
\end{subequations}
\\ \\
\textbf{\emph{Problem P2}}
\begin{subequations}\label{eq:robust_dwn}
\begin{align}
\underset{\substack{x^\downarrow_{ij}( k), x^{s\downarrow}_{j}(k)\\}}{\text{min}}\;\; & \sum_{i \in \mathcal{S}} \sum_{j \in \mathcal{D}} c_{ij}  \vert\, x^\downarrow _{ij}(k) \,\vert +  \sum_{j \in \mathcal{D}} c_{s}\, \vert\, x^{s\downarrow}_{j}(k)\,\vert &&\\
\text{subject to} \\
&  \sum_{j \in \mathcal{D}}x^\downarrow_{ij}(k) \leq \bar{s}_i^\uparrow(k)\,,  &&\forall i \in \mathcal{S},\\
&  \sum_{j \in \mathcal{D}}x^\downarrow_{ij}(k) \geq \bar{s}_i^\downarrow(k)\,,  &&\forall i \in \mathcal{S},\\
& \sum_{i \in \mathcal{S}}x^\downarrow_{ij}(k) + x^{s\downarrow}_{j}(k) = \bar{d}^\downarrow_j(k)\,,   &&\forall j \in \mathcal{D}\,.
\end{align}
\end{subequations}
\\
The terms $c_s$ in \eqref{eq:robust_up} and in \eqref{eq:robust_dwn} are introduced to weight the power transactions that cannot be satisfied by local sources. Slack variables $x_j^{s \uparrow}$ and $x_j^{s \downarrow}$ must be heavily penalized to discourage their use. In this purpose, we set   $c_s>c_{ij}, \; \forall i \in \mathcal{S}, \; \forall j \in \mathcal{D}$. It is worth noticing that all weights $c_{ij}$ and $c_s$ in \eqref{eq:robust_up} and \eqref{eq:robust_dwn} are strictly positive.
\\
\\
Note that in \textbf{\emph{Problem P1}} and in \textbf{\emph{Problem P2}} inequality constraints are introduced to properly bound the optimization variables, while the equality constraint represents the fulfilment of the overall demand at a given time instant.\\
Letting $x^{\uparrow *}_{ij}(k)$ and $x^{\downarrow *}_{ij}(k)$ be the optimal solutions of the two problems, the overall "average" transaction between source $i \in \mathcal{S}$ and sink $j \in \mathcal{D}$ is defined as
\begin{equation}\label{eq:reg_trans}
x^*_{ij}(k) \;=\;\frac{1}{2}(\vert x^{\uparrow *}_{ij}(k)\vert  + \vert x^{\downarrow *}_{ij}(k)\vert) \,.
\end{equation}
This value can be interpreted as a qualitative measurement of the importance of source $i$ in compensating power variations of  sink $j$. The higher this value is, the more important source $i$ is for balancing the variations of sink $j$, considering also the "distance" between them, captured by the weights $c_{ij}$.

\pagebreak
\subsection{Transactions projection onto shortest path}\label{subsec:trans_proj}
The optimal transactions  $x^*_{ij}(k)$ between sources and sinks must be now mapped onto the actual network graph. This operation serves to identify the importance of each edge, which will then determine which edges will be cut to create the network partitions.
\\
The transactions projection can be performed using different methods. For instance, the physical equations governing the network could be used to obtain the flows in each edge of the network, leading however to a complex and not scalable procedure. Here, it is proposed to project the optimal transaction between each source $i$ and sink $j$ on the shortest path connecting the two nodes. Other than being a general and easily-implementable approach, this has the further advantage of producing clusters that are as compact as possible.
Indeed, if the optimal transaction flow between a source and a sink has a high value, the two nodes will be probably included in the same network cluster together with edges of the shortest path connecting them, aiming to create a compact cluster.

Transactions must be mapped considering their orientation, since multiple transactions could cross the same edge with opposite directions. In fact, it is reminded that each transaction is directed from the source to the sink node by definition.

Consider the generic optimal transaction $x^*_{ij}(k)$, the shortest path between the source $i$ and the sink $j$ is expressed as the following ordered sequence of edges
\begin{equation*}
\mathcal{L}_{ij} = \left\lbrace[i,\gamma], [\gamma, \eta], [\eta, \sigma]...\,, [ \omega, j]\right\rbrace\,.
\end{equation*}
The projection of transaction  $x^*_{ij}(k)$ on the generic edge $e\in \mathcal{E}$ is denoted by the variable $x_{ij,e}(k)$. Since transactions are chosen to be projected on the associated shortest paths, it follows that  $x_{ij,e}(k)=0\,,\,\,\forall e \in \mathcal{E} \setminus \mathcal{L}_{ij}$. On the other hand, for each generic edge \mbox{$ e=[\alpha,\beta] \in \mathcal{L}_{ij}$}, the projection is performed according to the following convention
\begin{align}
\label{eq:proj_shortest_path}
x_{ij,e}(k) =
\left\{ 
\begin{aligned}
\;\;\;\;\,\hspace{0.2mm}x^*_{ij}(k) & \quad \textrm{if} \quad 
\alpha < \beta \\
\;\;- x^*_{ij}(k) & \quad \textrm{if} \quad 
\alpha > \beta 
\end{aligned} \right. \;\,.
\end{align}

\begin{figure}[t!]
	\includegraphics[width=0.8\linewidth]{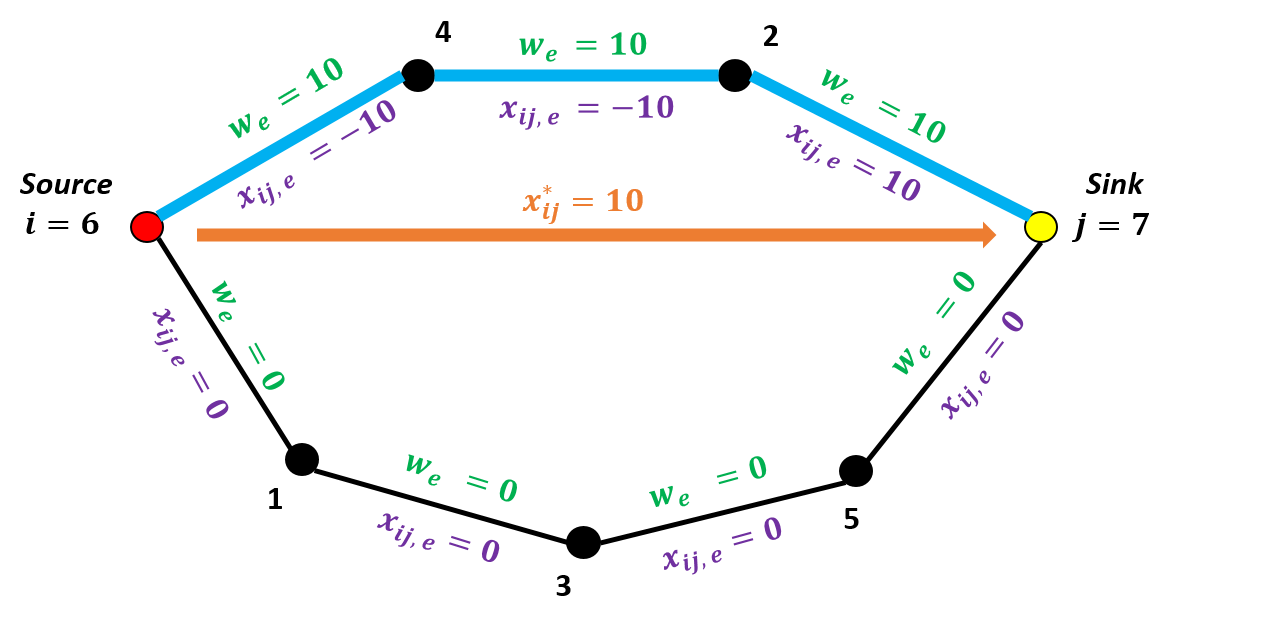}
	\centering
	\caption[Illustration of the projection onto the shortest path]{A simple illustration of the transaction projection onto the shortest path in the case of a single transaction.}
	\label{fig:illustration_projection}
\end{figure}

This allows to properly take into account different transactions, with the associated paths, which include the same edge but with opposite direction. 
This information is needed to determine the importance of edges considering the net amount of power flowing through it. Indeed, a new weight $w_e$ for each edge $e\in \mathcal{E}$ is now introduced, defined as the sum of all transactions' projections during the considered clustering period. It follows that
\begin{equation}
\label{eq:reg_resource_from_proj}
w_e = \sum_{k=0}^{N_c-1}\,\left\vert\sum_{i \in \mathcal{S}}\sum_{j \in \mathcal{D}} x_{ij,e}(k)\right\vert.
\end{equation}
The weights $w_e$ capture the relevance of each edge $e$ in releasing all power transactions taking place in the network over the clustering period.
\\
\\ 
Figure \ref{fig:illustration_projection} gives a simple illustration of the described procedure to map transactions on shortest paths. Node indices ranging from 1 to 7 are given by black numbers and each edge $e$ is characterized by a length $l_e=1$. The optimal transaction $x^*_{ij}=10$ between source $i=6$ and sink $j=7$ is depicted with an orange line (note that this is \textit{not} an edge of the graph). The shortest path connecting the source and the sink is indicated with a thick light blue line and it is given by the following edges: $\mathcal{L}_{ij} = \lbrace[6,4], [4,2], [2,7]\rbrace$. The projections $x_{ij,e}$ following equation \eqref{eq:proj_shortest_path} are indicated in purple while the final  $w_l$ terms, computed according to equation \eqref{eq:reg_resource_from_proj} are indicated in green. At this stage, consider edge $e = [4,2] \in \mathcal{L}_{ij}$. Since the first end-node of the edge has an higher index with respect to the second, it implies that $x_{ij,e} = -x^*_{ij} = -10$.

\subsection{Minimal edge-cut partitioning using \textsc{Metis}}\label{subsec:metis}

Our next goal is to partition the electric network into sub-networks so as to minimize the exchange of power between them. To this purpose, we perform a standard
\textit{k-way} partitioning of the graph, using the weights $w_e$ \cite{karypis2000multilevel}. The overall network is decomposed by minimizing the edge-cut, i.e. the sum of the weights of the edges connecting individual clusters.
Graph partitioning and clustering problems arise in many fields of science and technology and a wide literature is available, especially regarding the {k-way} partitioning methods.
Therefore, the already available partitioning tool \textsc{Metis} \cite{bib:METIS_doc} is used here, applying it to the network graph $\mathcal{G}(\mathcal{V},\mathcal{E})$ with the weights $w_e$. \\The partitioning procedure is carried out by fixing a-priori the number of $M$ clusters. In fact, determining both the size and the optimal number of clusters leads to a very complex problem formulation, and some heuristic approaches have been proposed in the literature  \cite{bib:Kiran2016}.
\\ \\
At the end of the proposed procedure, and in any clustering period, the network graph is decomposed into $M$ connected and non overlapping sub-graphs,  i.e. \mbox{$\mathcal{G} =\mathcal{G}_1\cup \dots \cup \mathcal{G}_{M}\,$}, where each sub-graph \mbox{$\mathcal{G}_{h}(\mathcal{V}_h,\mathcal{E}_h)$}, $h=1,\dots ,M$, includes a sub-set of dispatchable sources $\mathcal{S}_{h} \subseteq \mathcal{V}_h$, a sub-set of sinks $\mathcal{D}_{h} \subseteq \mathcal{V}_h$, and possibly a sub-set of storage units connected to the sources $\mathcal{B}_{h} \subseteq \mathcal{S}_h$.
Once partitions have been defined, both dispatchable sources and storage units must be coordinated by the two-layer control architecture described in the following, so as to ensure the continuous compensation of the sinks' power variations and the efficient exchange of power between the network partitions, if needed.

\section{Two-layer control of network clusters}\label{sec:two_layer_control}
The control algorithm is now specified according to the two-layer scheme shown in Figure \ref{fig:controlscheme}.
At each sampling time $k=0,\dots,N_c-1$ (for simplicity, the first clustering period is considered again), the proposed two-layer control architecture performs the following operations
\begin{itemize}
\item The local MPC regulator of each cluster  $h\in\{1,\hdots,M\}$ computes, in parallel to the others, the optimal output power variations of local dispatchable sources and batteries to compensate the load power variability. In case local sources are not sufficient to balance this variability, the MPC regulator sends to the upper supervisory layer a power request, denoted as $r^*_h(k)$. The remaining power availability in the local sources must be also communicated to the supervisory layer. 
\item The supervisory layer is activated just if $r^*_h(k) \neq 0$ for any generic cluster $h$, meaning that at least one cluster needs support. This upper layer is executed immediately after the local MPC regulators, within the same sampling time $\tau$. Through a properly defined optimization procedure, it computes the optimal variation of the output power of each cluster $h$, denoted as $\Delta y^*_h(k)$, to compensate the issued power requests. The power variation $\Delta y^*_h(k)$ is then sent to the local MPC cluster regulators, which execute this variation at the next time instant, i.e. $k+1$.
\end{itemize}

The specific control problems can be now stated. The local cluster MPC regulators are described first; then, the supervisory layer is discussed.

\subsection{Decentralized local MPC cluster regulator}

Consider a cluster $h\in\{1,\dots,M\}$, and any sampling time $k=0,\dots,N_c-1$. The control actions are computed as the solution of a suitable constrained optimization problem. Let \mbox{$S_h=|\mathcal{S}_h|$} be the number of dispatchable sources and let $s_{hi}$, $i=1,\dots,S_h$, be the corresponding output power variation with respect to a given nominal profile. Moreover, define by $B_h=|\mathcal{B}_h|$ the number of storage elements available and by $e_{hi}$, $i=1,\dots,B_h$, the corresponding variation of stored energy, modeled as a pure integrator
\begin{align}\label{batt}
e_{hi}(\tilde{k}+1)=e_{hi}(\tilde{k})\,-\,\tau\,b_{hi}(\tilde{k}),\,\qquad \tilde{k}=0,\dots,N_c-1,
\end{align}
where $b_{hi}(\tilde{k})$ indicates the output power variation of the storage unit at the generic time instant $\tilde{k}$.\linebreak
The dispatchable sources' output power variations $s_{hi}$, the variation of stored energy in batteries $e_{hi}$ and the effective output power variation $b_{hi}$ must instantaneously respect the following bounds  for \mbox{$\tilde{k}=0,\dots,N_c-1$}
\begin{align}
&\bar{s}^\downarrow_{hi}(\tilde{k}) \,\leq \,s_{hi}(\tilde{k})\,\leq \,\bar{s}^\uparrow_{hi}(\tilde{k})\,, \qquad \forall \,i = 1,\dots,S_h,\\
&\bar{b}^\downarrow_{hi}(\tilde{k}) \,\leq \,b_{hi}(\tilde{k})\,\leq \,\bar{b}^\uparrow_{hi}(\tilde{k})\,, \qquad \forall \,i = 1,\dots,B_h,\\
&\bar{e}^\downarrow_{hi}(\tilde{k}) \,\leq \,e_{hi}(\tilde{k})\,\leq \,\bar{e}^\uparrow_{hi}(\tilde{k})\,, \qquad \forall \,i = 1,\dots, B_h\,.
\end{align}
The net output power variation of each cluster $h$ computed through the power balance equation
\begin{align}\label{bilancio}
y_h(\tilde{k})\,=\, \sum_{i=1}^{S_h}s_{hi}(\tilde{k})\;+\;\sum_{i=1}^{B_h}b_{hi}(\tilde{k})\;-\sum_{j=1}^{D_h}d_{hj}(\tilde{k})\,,
\end{align}
where $d_{hj}$ is the power variation of load $j$ included in cluster $h$.\\\\
The requirements that the cluster is balanced and self-sufficient correspond to the constraint $
y_h(\tilde{k})\;=\;0\,$.
However, to avoid infeasibility issues, when local dispatchable sources and storage units are not able to balance sinks' power variations inside the cluster, a slack variable $r_h(\tilde{k})$ is introduced, so that the previous constraint becomes
\begin{align}\label{bilancio1}
y_h(\tilde{k})\,+\, r_h(\tilde{k})\,=\,0\,.
\end{align}
The term $r_h(\tilde{k})$ will be highly penalized in the cost function, so that an optimal solution $r_h^*(\tilde{k})\neq 0$ is computed only if cluster $h$ is unable to balance the local demand and requires power support from other clusters, called \textit{supporting} clusters. In this case, the upper distributed supervisory layer will be activated to enforce the supporting clusters to modify their output power. To this purpose, a variable $\Delta y_h^*(\tilde{k}-1)$ is introduced, acting as an output power offset committed by the supervisory layer at the previous time instant $\tilde{k}-1$. Therefore, constraint \eqref{bilancio1} is reformulated as follows
\begin{align}\label{eq:net_output}
y_h(\tilde{k})\,+\, r_h(\tilde{k})\,=\,\Delta y^*_h(\tilde{k}-1)\,.
\end{align}
To motivate the time shift in \eqref{eq:net_output}, note that we assume that the supervisory layer is activated right after the local MPC regulator within the same sampling time; however, its solution is available to the local MPC regulators at the next time instant.\\
Finally, the availability of power reserve in each cluster is also modeled, since this information will be exploited by the supervisory layer. For the sake of simplicity, only the available reserves in dispatchable sources are considered, while  storages are assumed to be used only for local compensation\footnote{Reserves from storage units can be  easily included, but their definition requires a more detailed formulation considering both the capability limits and the stored energy. For more details, the reader is referred to \cite{8848405}. }. Therefore, the upward and the downward reserves of cluster $h$, denoted as $\Delta s^{\uparrow}_h$ and $\Delta s^{\downarrow}_h$, respectively, are  defined as

\begin{align}\label{eq:reserves1}
\Delta s^{\uparrow}_h(\tilde{k})\,&=\,\sum_{i=1}^{S_h} \Big ( \bar{s}^{\uparrow}_{hi}(\tilde{k}) \,-\,{s}_{hi}(\tilde{k}) \Big )\,,\\
\Delta s^{\downarrow}_h(\tilde{k})\,&=\,\sum_{i=1}^{S_h} \Big ( \bar{s}^{\downarrow}_{hi}(\tilde{k}) \,-\,{s}_{hi}(\tilde{k})\Big )\,\label{eq:reserves2}
\end{align}
At this stage, the cost function of the cluster MPC regulator can be defined. To this end, let the positive integer $N_p$ be the adopted prediction horizon, for simplicity considered to be equal for all the clusters. Moreover, assume to be at the generic time instant $k\in\{0,\hdots,N_c-1\}$, and define $\bar{N}_p(k)=min(N_p,N_c-k)$. Then, the cost function to be minimized is defined as
\begin{align}
J_h(k)\,=\, \sum_{\varphi=k}^{\bar{N}_p(k)-1} \Bigg [ \sum_{i=1}^{S_h} \gamma_s \,{s}^2_{hi}(\varphi) + \sum_{i=1}^{B_h} \gamma_b\, {b}^2_{hi}(\varphi) +  \gamma_r \,  r_h^2(\varphi) \Bigg ]\,,
\end{align}
where $ \gamma_r,\gamma_s,\gamma_b$ are positive constants and $ \gamma_r \gg \max\{\gamma_s,\gamma_b\}$ to strongly penalize the use of the slack variable $ r_h$.
\\
\\
Defining
\begin{align}
\vec{s}_h(k)&=[s_{h1}(k),\dots,s_{hS_h}(k),\dots,s_{h1}(\bar{N}_p(k)-1),\dots,s_{hS_h}(\bar{N}_p(k)-1)],\\
\vec{b}_h(k)&=[b_{h1}(k),\dots,s_{hB_h}(k),\dots,b_{h1}(\bar{N}_p(k)-1),\dots,b_{hB_h}(\bar{N}_p(k)-1)],\\
\vec{r}_h(k)&=[r_h(k),\dots,r_h(\bar{N}_p(k)-1)],
\end{align}
the local MPC optimization problem can now be stated as follows

\begin{subequations}
\begin{align}
\min\limits_{\substack{
		\vec{s}_{ h }(k)\; 	\vec{b}_{ h }(k) ,\vec{r}_{h}(k)} }\,& \qquad \quad J_{h} (k) \\
\text{subject to} \quad & \nonumber\\
& \text{\eqref{batt}-\eqref{bilancio}, \eqref{eq:reserves1}-\eqref{eq:reserves2}}, \quad \quad    \quad \; \quad \quad \quad \quad\tilde{k}=k,\dots,\bar{N}_p(k)-1,\\
& y_h(\tilde{k})\,+\, r_h(\tilde{k})\,=\,\Delta y^*_h({k}-1), \quad \quad    \quad \tilde{k}=k,\dots,\bar{N}_p(k)-1, \label{eq:net_output1}\\
&e_{hi}(N_c) \,=\, e_{hi}(0),\; \; \forall \,i \in \mathcal{B}_h\,, \qquad \; \; \,\quad \text{if $\bar{N}_p(k)=N_c-k$}\,. \label{endconstr}
\end{align} \label{eq:localMPC}
\end{subequations}

Concerning constraints and \eqref{eq:net_output1}, \eqref{endconstr} some comments are in order. First, in \eqref{eq:net_output1} the right hand side of the equation is kept constant at the value computed at the previous time instant, being unknown the future behavior of the supervisor computing $\Delta y^*_h$. Second, constraint \eqref{endconstr} is included to guarantee that the resources stored at the beginning of the clustering period are recovered at its end, so that a complete discharge of the batteries is prevented and a sufficient amount of energy is available for the next clustering period.\\
\\
Denoting with the superscript ($^*$) the optimal value of the optimization variables and of the related quantities, at any time $k$ only the optimal values $\vec{s}_{h }^*(k)$, $\vec{b}_{h}^*(k)$, $\vec{r}_{h}^*(k)$ are implemented and the overall procedure is repeated at any new sampling time.\\
The optimal values of $r_h^{*}(k)$, $\Delta s^{\uparrow\,*}_h(k)$, and $\Delta s^{\downarrow\,*}_h(k)$ are also transmitted to the supervisory layer. If $r_h^{*}(k)=0$, the cluster $h$ does not require any additional contribution from the other clusters and can provide them with its reserves specified by $\Delta s^{\uparrow\,*}_h(k)$, and $\Delta s^{\downarrow\,*}_h(k)$. On the contrary, if If $r_h^{*}(k) \neq 0$, the supervisor must redirect power from the supporting clusters to cluster $h$.
%
%
\begin{rmk}
The adopted strategy corresponds to a standard Receding Horizon control approach whenever $\bar{N}_p=N_p$, while it is based on a shrinking horizon strategy when the current time index approximates the end of the clustering period. This is due to the fact that, when the clustering period ends, the structure of the clusters can change and, accordingly, the overall control algorithm must be reset.
\end{rmk}
\subsection{Distributed clusters' supervisory control}
The supervisory layer is based on a fully distributed algorithm, where each agent is implemented on the top of a network cluster and directly interacts with its neighbors based on a pre-defined communication graph.
For the sake of clarity, before describing in detail the adopted distributed algorithm, the centralized formulation of the supervisory problem is presented.\\
As shown in the previous section, each local MPC regulator can act on the variable $r^{*}_h(k)$, taking values different from zero just in case internal sources and storage units of cluster $h$ are not sufficient to balance local sinks' power variations.
To overcome this issue, the supervisory layer commits the other clusters to vary their output net flow to support cluster $h$ by optimally varying the output power flows of the remaining clusters, i.e. selecting the variables $\Delta y_j(k)$, $\forall j\in\{1,\dots,M\} \setminus h$. First of all, the following constraint must be fulfilled
\begin{align}
\label{eq:sup_power_balance}
\sum_{h=1}^{M} \Delta y_h(k) \;=\;\sum_{h=1}^{M} r^*_h(k)\,,
\end{align}
to ensure the overall balance between the power requests by the local MPC regulators and the committed output variations by the supervisory layer.
From \eqref{eq:sup_power_balance}, it is evident that, in case two local MPC regulators send two opposite requests with equal magnitude, the overall request is null. This is done on purpose since the main objective is that the overall network is self-balanced, even though the single clusters are not.
The committed output variation must respect the upward and downward power reserve available in each cluster, i.e.
\begin{align}
\label{eq:sup_reserves}
\qquad \Delta s^{\downarrow*}_h(k)\, \leq \, \Delta y_h(k) \, \leq \,  \Delta s^{\uparrow*}_h(k) \,, \qquad \forall h \in\{1,\dots,M\} \;.
\end{align}

At this stage, the centralized supervisory problem is stated as follows
\begin{align}\label{eq:sup_problem}
\begin{split}
\min\limits_{\Delta y_{h}(t)} \quad\sum_{h=1}^{M} c_h\,\Delta y^2_h(t)& \\[2mm]
\text{subject to} \qquad
\eqref{eq:sup_power_balance} \,\text{-}\,  \eqref{eq:sup_reserves}&
\end{split}
\end{align}

where the terms $c_{h}>0$ can be different among the clusters. As the supervisory layer solves \eqref{eq:sup_problem}, the optimal values $ \Delta y^*_{h}(t)$ are obtained, and these are communicated to local MPC regulators so as to be executed at the next time instant, as described by the constraint \eqref{eq:net_output}.\\
We notice that problem \eqref{eq:sup_problem} is constituted by a separable cost function, some local constraints, i.e.  \eqref{eq:sup_reserves}, and by a power balance constraint which couples clusters' variables, i.e. \eqref{eq:sup_power_balance}. The optimization problem  \eqref{eq:sup_problem} can be also reformulated in the following  form
\begin{subequations}\label{eq:SUP}
\begin{align}
\min_{x_1,\,\dots,\,x_{M}} \;\;\;\; &\sum_{h=1}^{M} f_h(\,x_h\,)  \\[1mm]
\text{subject to} \quad
&\,x_h \in X_h\,, \quad \forall k \in \{1,\dots,M\}\,,\\
&\sum_{h=1}^{M} E_{h} x_h = q\;, \label{eq:coupl_constr}
\end{align}
\end{subequations}
where $x_h=\Delta y_h(t)$, $f_h(x_h)=\Delta y^2_h(t)$,  $X_h$ is a polyhedral set defined by constraints \eqref{eq:sup_reserves},  while  $\sum_{k=1}^{M} E_{h} x_h = q$ represents the coupling constraint \eqref{eq:sup_power_balance}.
The optimization problem \eqref{eq:SUP}, here named \textit{primal problem}, is a  convex optimization problem, which can be easily solved through distributed optimization theory using Lagrangian Relaxation, see \cite{bertsekas1997nonlinear}. Precisely,
a fully distributed approach is adopted in this work, meaning that each agent can directly interact with the others through a communication graph, without the need of any central coordination entity  \cite{bertsekas1997nonlinear}.
An overview of this algorithm is given in the next Section.

\subsubsection{Distributed Consensus ADMM}
The adopted distributed approach is named Dual Consensus ADMM (DC-ADMM) \cite{chang2014multi,chang2016proximal,banjac2019decentralized}.
As mentioned, the approach is based on a direct interaction  among the different agents, so as to solve the \textit{primal problem} in \eqref{eq:SUP} in a distributed fashion.
First, a multi-agent communication network  is introduced, modeled as an undirected graph \mbox{$\mathcal{{G}}^c=\{\mathcal{{V}}^c,\mathcal{{E}}^c\}$}, where \mbox{$\mathcal{{V}}^c=\{1,\dots,M\}$} is the set of nodes (i.e. the distributed supervisors of each cluster) and $\mathcal{{E}}^c$ is the set of edges  (i.e. the communication links).
If agent $i$ and agent $j$ can directly exchange messages, they are named \textit{neighbors} and therefore  $(i,j) \in \mathcal{E}^c$. Thus,  for each agent $i$, the subset $\mathcal{N}_i=\{j\in \mathcal{V}^c \, | \, (i,j) \in \mathcal{E}^c\}$ including all neighbors of $i$ can be defined, together with the parameter $n_i=\vert \mathcal{N}_i \vert $ denoting their number.

\pagebreak
The DC-ADMM algorithm relies on an iterative procedure which asymptotically converges to the optimal solution  if the  communication graph  $\mathcal{G}^c=\{\mathcal{V}^c,\mathcal{E}^c\}$ is connected and the primal problem is convex, see \cite[Th. 2]{chang2014multi}.
We note that our setup matches these assumptions.
Duality optimization theory plays a central role in the definition of the DC-ADMM method. To this purpose, the Lagrangian function of \eqref{eq:SUP} is introduced
\begin{align*}
\begin{split}
L(x_1,\dots,x_{M}, \lambda) \,&= \, \sum_{h=1}^{M} f_h(\,x_h\,) \,+\, \lambda \Bigg (\sum_{h=1}^{M} E_{h} x_h - q \Bigg ) \, = \, \sum_{h=1}^{M}\Big \{\; f_h(\,x_h\,) \;+ \; \lambda \,E_{h} \,x_h \;- \;\lambda \frac{q}{M}\;\Big \}\,,
\end{split}
\end{align*}
which is obtained through the relaxation of the coupling constraint  \eqref{eq:coupl_constr} and the introduction of the dual variable $\lambda$.
Therefore, the dual problem of \eqref{eq:SUP} can be stated as

\begin{align}\label{eq:dual_problem}
\max_{\lambda}\;\sum_{h=1}^{M}\min_{x_h \in X_h }\Big \{\; f_h(\,x_h\,) \,+\, \lambda \,E_{h} \,x_h \;- \;\lambda \frac{q}{M}\;\Big \}\,,
\end{align}
which can be also written as

\begin{align} \label{eq:dual_prob_reduced}
\min_{\lambda}\;\;\sum_{h=1}^{M}\Bigg[\,\lambda \frac{q}{M}\;\, \underbrace{- \min_{x_h \in X_h }\Big \{\; f_h(\,x_h\,) \;+ \; \lambda \,E_{h} \,x_h\;\Big \}}_{\phi_h(\lambda)}\,\Bigg].
\end{align}
Since the communication graph among agents is connected, problem \eqref{eq:dual_prob_reduced}  is equivalent to
\begin{subequations}\label{eq:dual_prob_reduced_cons}
\begin{align}
\underset{\lambda_1,\,\dots,\,\lambda_{M} \label{eq:dual_cons_cost_function}}{\min}\quad & \sum_{h=1}^{M}\Big (\,\lambda_h \frac{q}{M}\, + \,{\phi_h(\lambda_h)}\,\Big )\\
\text{subject to} \quad & \lambda_h\,=\, \lambda_j,\, \qquad \forall h \in \mathcal{V},\;\; \forall j \in \mathcal{N}_h\,.\label{eq:dual_cons_constr}
\end{align}
\end{subequations}

At this stage, the cost function \eqref{eq:dual_cons_cost_function} can be now split among agents, where each of them can solve a local problem  with respect to a local copy of the dual variable $\lambda$, denoted as $\lambda_h$.
Nevertheless, at convergence, all agents must reach a consensus on the local copies of $\lambda$, so that the coupling constraints \eqref{eq:dual_cons_constr} are respected for all neighboring agents.
As described in \cite{mateos2010distributed}, the final step to derive the DC-ADMM algorithm consists in using the standard Consensus ADMM (C-ADMM) to solve the dual problem  \eqref{eq:dual_prob_reduced_cons}.
The mathematical steps required to perform this operation are here omitted, and they are reported in \cite[Section IV-A]{chang2014multi}.
The final form of the DC-ADMM method is given in Algorithm \ref{alg:DC_ADMM}. 
As evident, all agents act completely in parallel, solving in sequence Steps 6-9 at each iteration. At each iteration $i$, agent $h$ receives the optimal values of the local dual variables' of its neighbors computed at the previous iteration, i.e. $\lambda^{i-1}_j$ $ \forall j\in \mathcal{N}_h$, 
computes the optimal value of the local optimization variable $x_h^i$ at Step 7 and then updates local dual variable $\lambda^{i}_h$ in Step 8. Finally, an additional auxiliary variable $p_h^i$ is updated in Step 9, based on the difference between the copies of the local dual variables agent $h$ and its neighbors. The variable $p_h^i$ serves in fact to guarantee convergence of the dual variables' copies to the same value.

\medskip
As shown in the numerical experiments (Section \ref{sec:num_res}), the supervisory layer, based on the DC-ADMM algorithm, manages to find the optimal solution of \eqref{eq:sup_problem} in few iterations.
In fact, the supervisory layer solves a simple static problem, for computing the optimal power exchanges of resource among the different network clusters.
\begin{algorithm}[!h]
\begin{algorithmic}[1]
	\State{Select $c>0$ as a tuning parameter}
	\State $\lambda_h^0 = 0, \,\,  p_h^0 = 0 \quad \forall \,h \in \mathcal{V}^c$
	\State $i = 1$
	\Repeat
	\For{$\forall \,h \in \mathcal{V}^c$, in parallel,}	
	\State  agent $h$ receives from its neighbors $\lambda^{i-1}_j$ with $j \in \mathcal{N}_h$
	\State \begin{flalign*}
	{x}_h^{\,i} =  \,&\underset{\mathbf{x}_h \in X_h}{\text{argmin}}  \Bigg\lbrace f_h(\,{x}_h\,) + \frac{c}{4 \, n_h}\Bigg\Vert \,\frac{1}{c}\left({E}_h\,{x_h} - \frac{q}{M}\right)\, -\, \frac{1}{c}{p}_h^{i-1} \,+\sum_{\forall j\in\mathcal{N}_h}\left( \lambda^{i-1}_h - \lambda^{i-1}_j \right)\Bigg\Vert_2^2\;\Bigg\rbrace && 			\end{flalign*}
	\State
	\begin{flalign*}
	{\lambda}^{i}_h=\; &\frac{1}{2\,n_h}\Bigg(\sum_{\forall j\in\mathcal{N}_h}\big(\lambda^{i-1}_h - \lambda^{i-1}_j\big) - \frac{1}{c}{p}_h^{i-1} +  \frac{1}{c}\big({E}_h{x}_h^{\,i} - \frac{1}{M}{q}\big) \Bigg) &&
	\end{flalign*}
	\State \begin{flalign*} {p}^{i}_h =\; p^{i-1}_h + c\sum_{ \forall j\in\mathcal{N}_h}\left(\lambda^{i}_h - \lambda^{i}_j \right) && \end{flalign*}\label{step:dualdual_update_dcadmm}
	\EndFor
	\State $i \leftarrow i + 1$
	\Until a predefined stopping criterion is satisfied.
\end{algorithmic}
\caption{ DC-ADMM for solving \eqref{eq:sup_problem}}
\label{alg:DC_ADMM}
\end{algorithm}

\medskip
To summarize, the fundamental operations of the described two-layer architecture  are represented in Algorithm \ref{alg:two_layer_control}, considering just the first clustering period for simplicity.
\begin{algorithm}[!h]
\begin{algorithmic}[1]
	\State{Initialize $\Delta y_h^*(-1)=0$}\vspace{1mm}
	\For{ $k\in \{0,\hdots,N_c-1\} $} \vspace{2mm}
	\State  The \textbf{local MPC regulator} of each cluster $h\in \{1, \hdots, M\}$ \textbf{does}\vspace{1mm}
	\State \hspace{0.5cm} Measure load power variation $d_{hj}(k)$, $j \in \mathcal{D}_h$, and the power variation  $\Delta y_h^*(k-1)$ \vspace{1mm}
	\State \hspace{0.5cm} Solve \eqref{eq:localMPC} and implement $s^*_{hi}(k)$ and $b^*_{hj}(k)$, with $i \in \mathcal{S}_h$ and $j \in \mathcal{B}_h$ \vspace{1mm}
	\State \hspace{0.5cm} Send the request $r_h^*(k)$ and the reserves $\Delta s^{\uparrow\,*}_h(k), \Delta s^{\downarrow\,*}_h(k)$ to the supervisory layer  \vspace{2mm}
	\State  \textbf{if} $ \exists \, r_h^*(k) \neq 0$, the \textbf{distributed supervisor} of each cluster $h\in \{1, \hdots, M\}$  \textbf{does}  \vspace{1mm}
	\State \hspace{0.5cm} Receive $r_h^*(k)$, $\Delta s^{\uparrow\,*}_h(k), \Delta s^{\downarrow\,*}_h(k)$ \vspace{1mm}
	\State \hspace{0.5cm} Execute Algorithm \ref{alg:DC_ADMM} in cooperation with the other clusters' supervisors \vspace{1mm}
	\State \hspace{0.5cm} Send to the local MPC regulator of cluster $h$ the optimal power variation $\Delta y_h^*(k)$ \vspace{1mm}
	\EndFor
\end{algorithmic}
\caption{ \;Two-layer control architecture operations}
\label{alg:two_layer_control}
\end{algorithm}

\pagebreak
\section{Numerical results}\label{sec:num_res}
\begin{figure}[b!]
\centering
\includegraphics[width=0.6\linewidth]{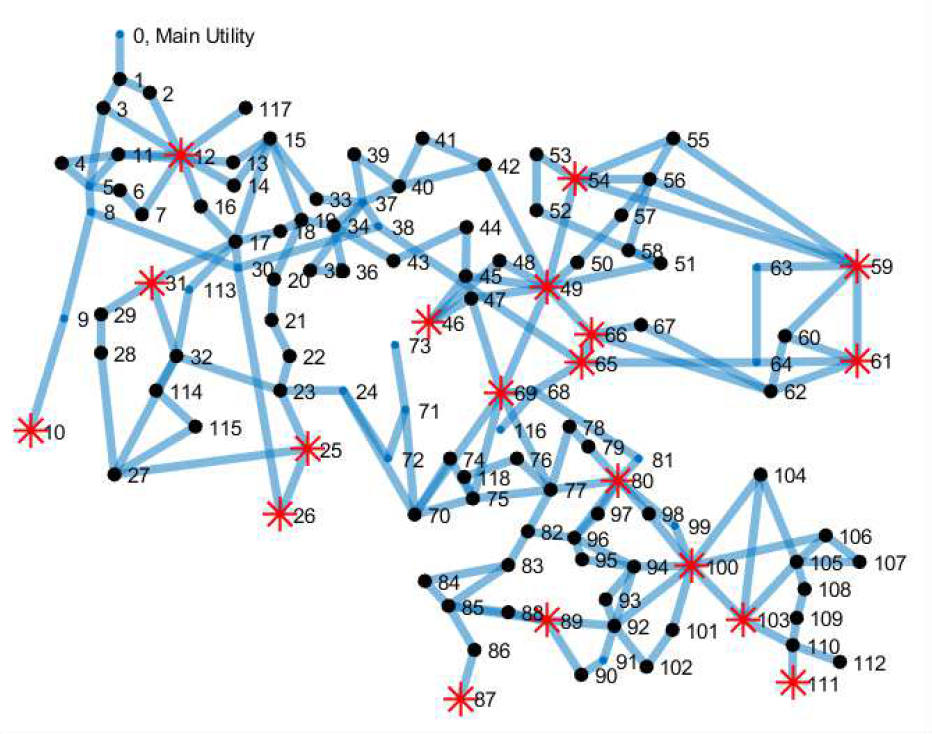}
\caption{IEEE 118 bus-system: Source nodes (red stars), Sink nodes (black dots).}
\label{fig:ieee118}
\end{figure}

The overall proposed architecture has been tested to control a large-scale electrical network, i.e. the IEEE 118-bus system, whose  data are reported in \cite{bib:IEEE_118_src}.
\\
A schematic of the network graph is depicted in Figure \ref{fig:ieee118}, showing the position of sources (i.e. generators) and sinks (i.e. loads). Moreover, it is assumed a storage element (i.e. a battery)  is present at each source node.
The network is connected to the main utility through the node 0, acting as the slack node.

\begin{figure}[t!]
\centering
\begin{tabular}{cc}
	\includegraphics[width=0.45\linewidth]{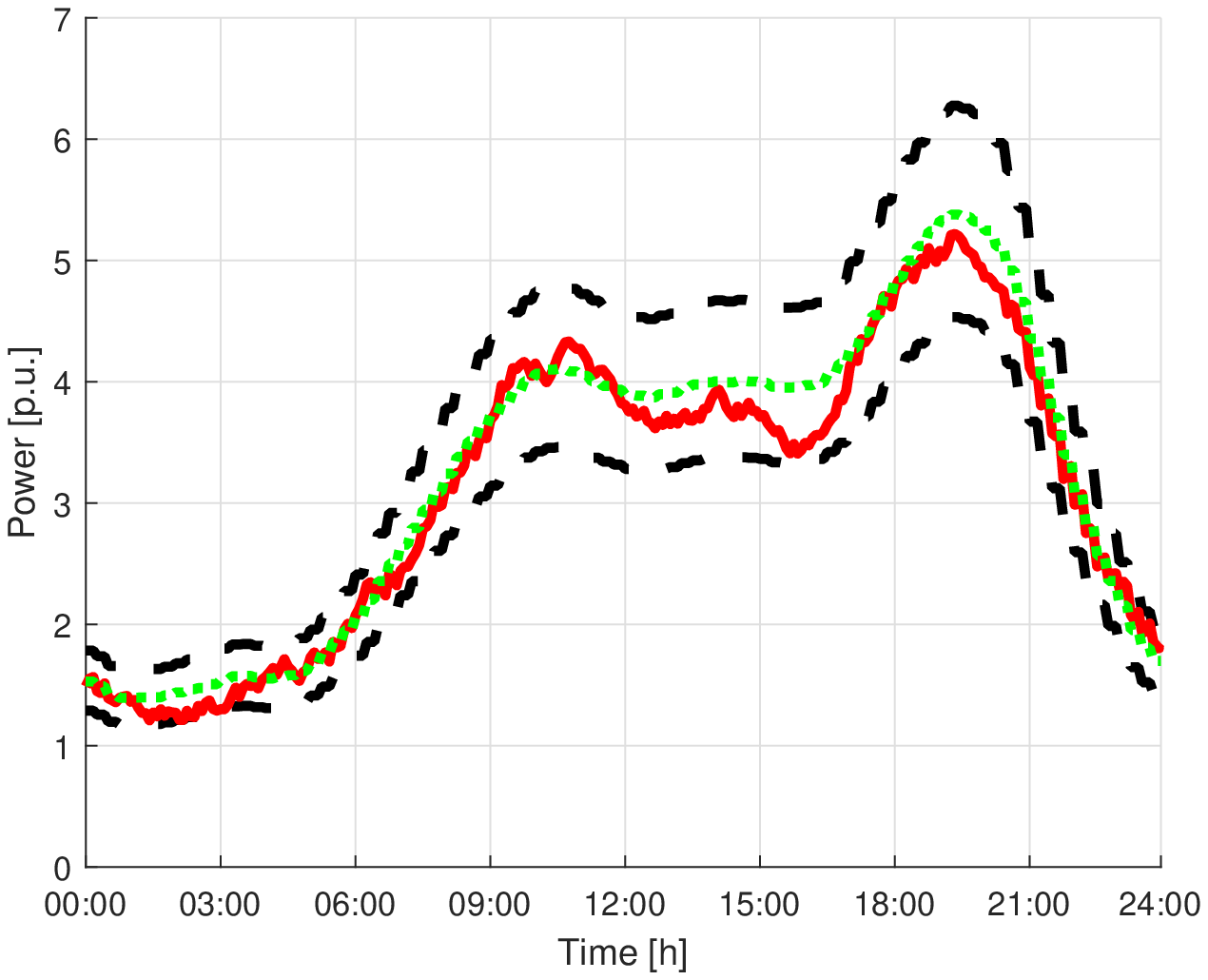} & \includegraphics[width=0.45\linewidth]{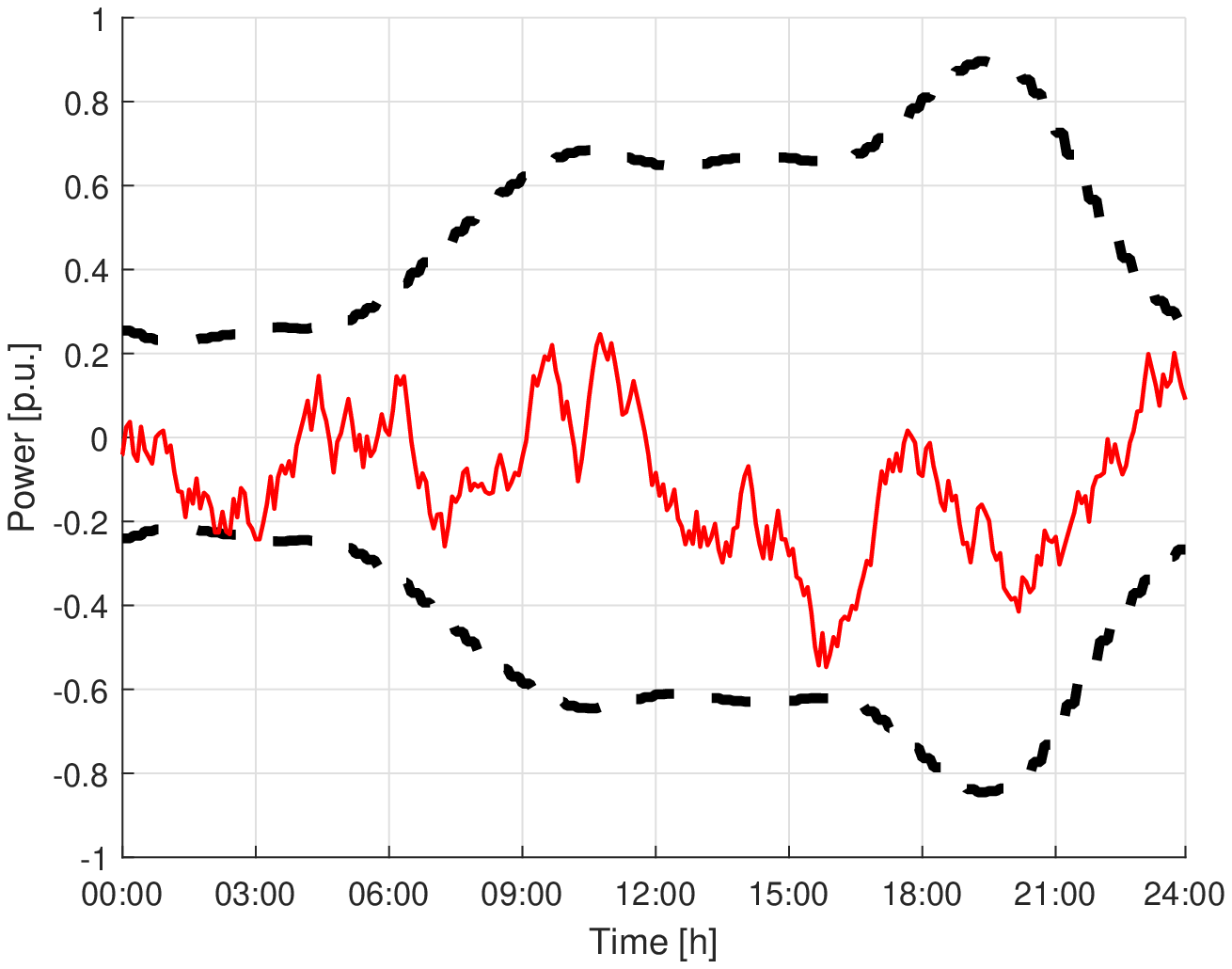}\\
	(a) & (b)
\end{tabular}
\caption{Sink at node 1: (a)  forecast/nominal demand (dotted green), worst-case maximum and minimum demand (dashed black), effective demand (solid red); (b) effective demand variability (solid red), worst-case maximum and minimum demand variability (dashed black).}	\vspace*{-5mm}
\label{fig:load}
\end{figure}

\begin{figure}[t!]
\centering
\begin{tabular}{cc}
	\includegraphics[width=0.45\linewidth]{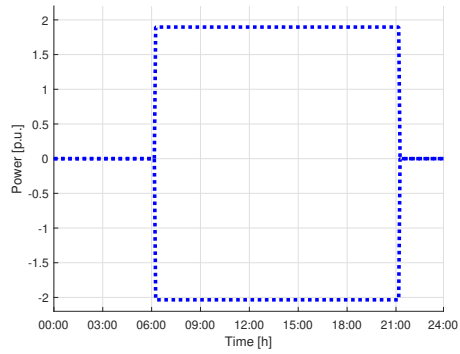}
\end{tabular}
\caption{Capability limits of source at node 25, expressed as difference between the absolute limits and the pre-scheduled production.}
\label{fig:gen}
\end{figure}

Consistently with real operation of electrical grids, it is assumed that the nominal power flows of the network have been already determined through an optimal power flow procedure, for instance performed on a day-ahead basis as in \cite{8848405}. However, during the real-time operation, sinks' power demand varies with respect to the nominal forecasts, causing a variation of the pre-scheduled power exchange with the main utility if not promptly compensated. Figure \ref{fig:load}(a) reports the actual behavior of the sink at node 1, where it is evident that the effective demand differ from the forecasted/nominal one, even though it is still contained in some worst-case bounds. Figure \ref{fig:load}(b) shows the effective sink's demand expressed as variation with respect to the nominal values, together with the worst-case bounds, computed as in \eqref{eq:up_demand}-\eqref{eq:down_demand}.
\\
Figure \ref{fig:gen} shows the capability limits of the dispatchable source at node 25, also expressed as variation with respect to the nominal profile, see \eqref{eq:up_gen}-\eqref{eq:down_gen}.
In this case, the source can be exploited to compensate the external sinks' demand variations just between 06:00 and 21:00. The rest of sources and sinks are characterized by different but similar features.

\begin{figure}[h!]
\centering
\includegraphics[width=0.45\linewidth]{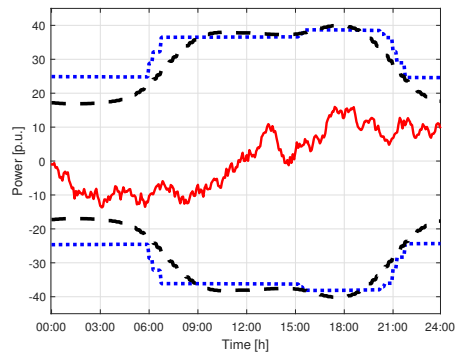}
\caption{Overall network: total sources upward/downward differential capability (dotted blue),  total sinks' worst-case (dashed black) and  effective variability (solid red).}
\label{fig:ieee118_reserves}
\end{figure}

Considering the whole network, the total capability of sources, the total worst-case bounds and effective demand variability of sinks are reported in Figure \ref{fig:ieee118_reserves}.
If dispatchable sources and batteries were not controlled to compensate the sinks' demand variability, the power exchange with the main utility would be affected by major fluctuations around the nominal power profile.

\begin{figure}[h]
\centering
\begin{tabular}{cc}
	\includegraphics[width=0.45 \linewidth]{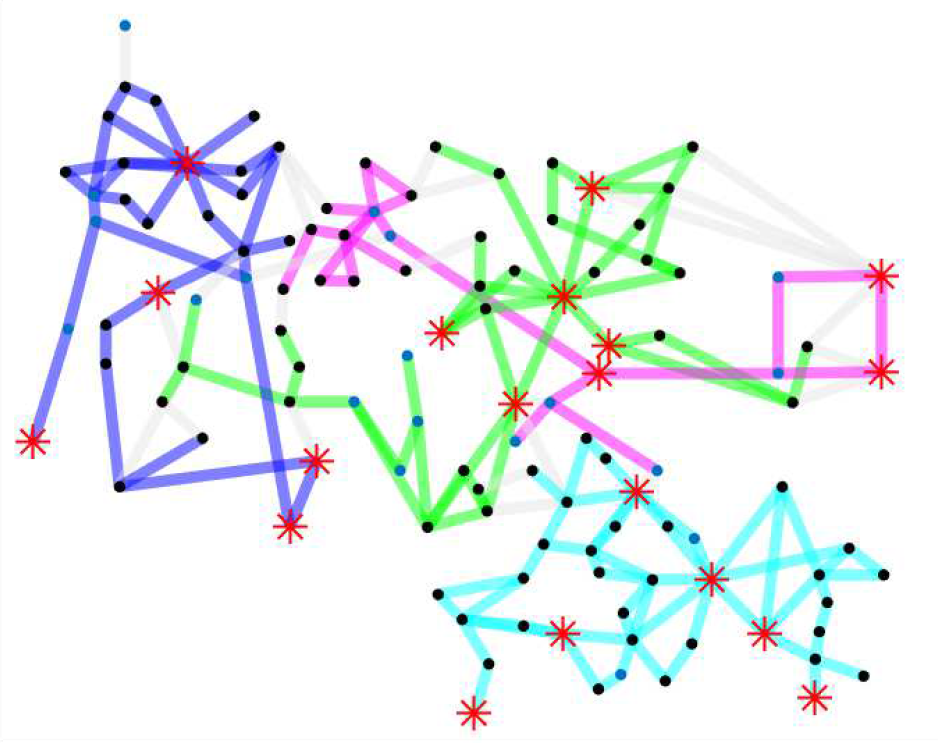} & \includegraphics[width=0.45\linewidth]{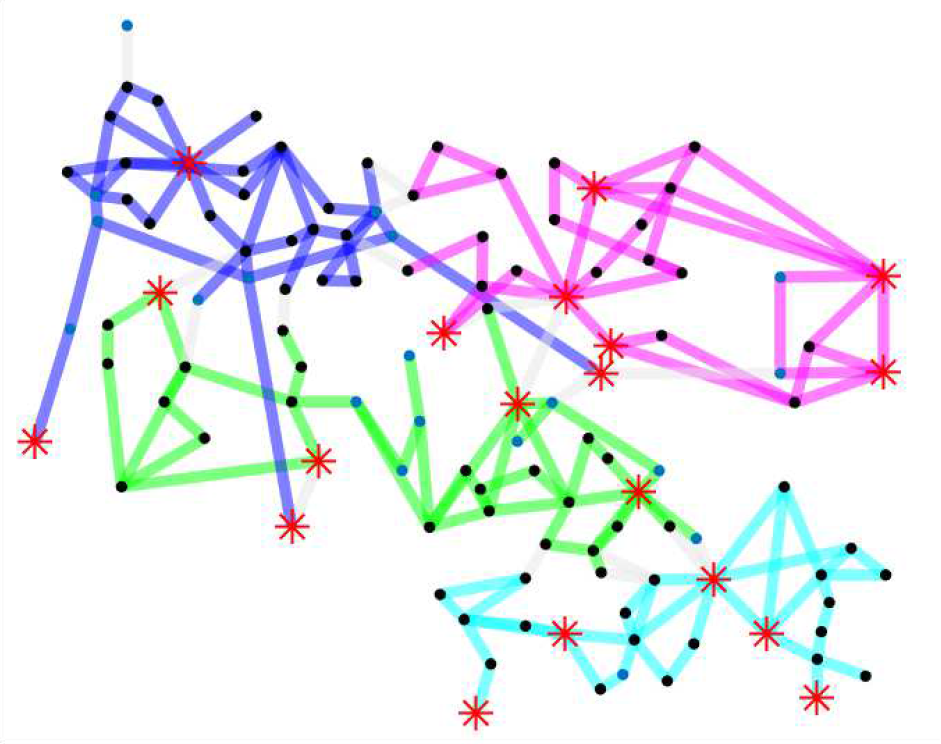}\\ 	(a) & (b)  \\
	\includegraphics[width=0.45\linewidth]{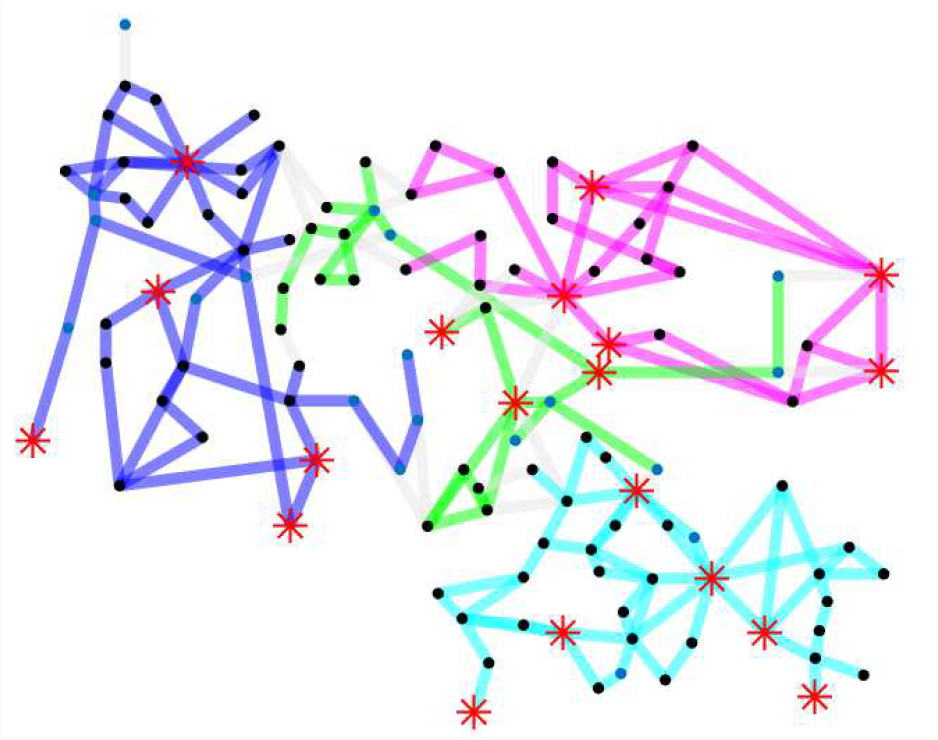} & \includegraphics[width=0.45\linewidth]{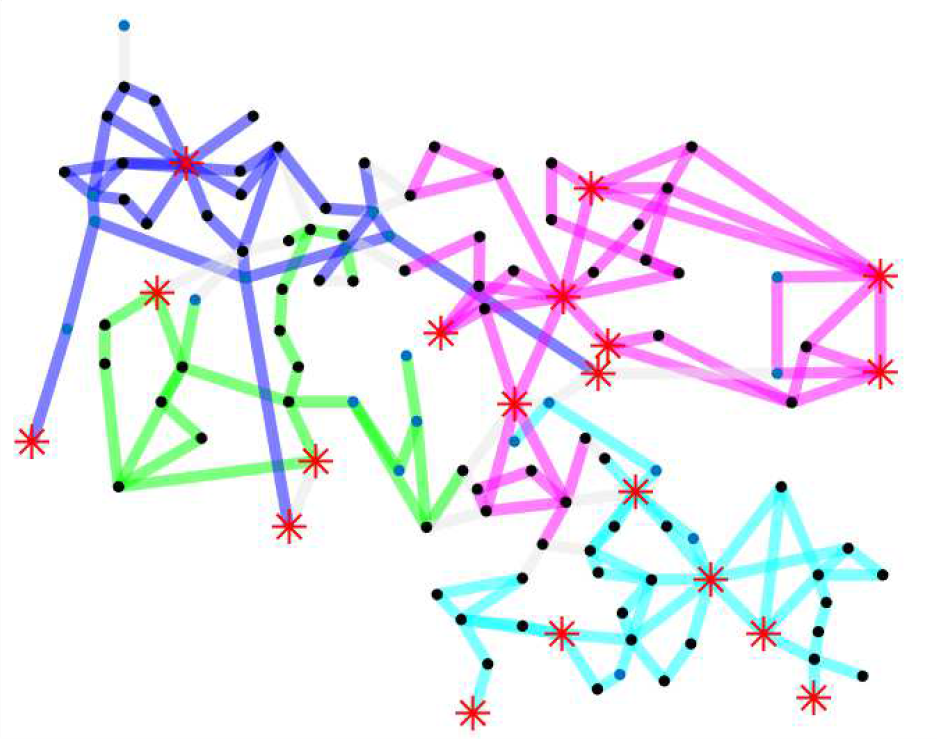}\\	(c) & (d)
\end{tabular}
\caption{Network clustering between (a) 00:00 and 06:00, (b) between 06:00 and 12:00, (c) between 12:00 and 18:00, and (d) between 18:00 and 24:00: Cluster 1 (green), Cluster 2 (cyan), Cluster 3 (blue) and Cluster 4 (magenta).}
\label{fig:clustering}
\end{figure}

\begin{figure}[h]
\centering
\begin{tabular}{cc}
	\includegraphics[width=0.45\linewidth]{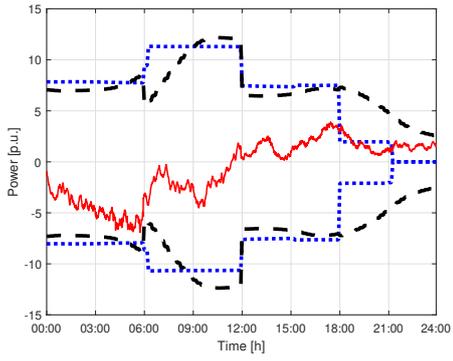} & \includegraphics[width=0.45\linewidth]{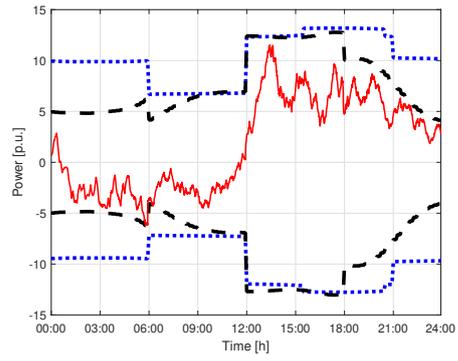}
	\\  (a) & (b)  \\
	\includegraphics[width=0.45\linewidth]{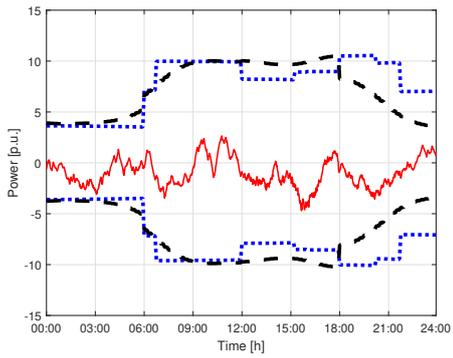} & \includegraphics[width=0.45\linewidth]{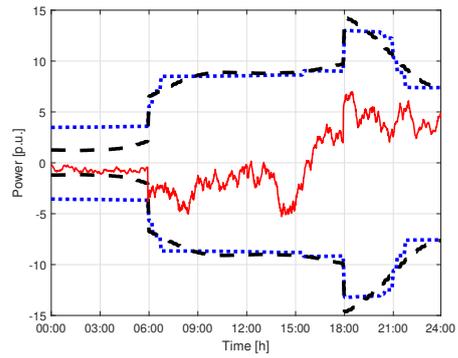}
	\\	(c) & (d)
\end{tabular}
\caption{(a) Cluster 1, (b) Cluster 2, (c) Cluster 3 and (d) Cluster 4: total sink upward/downward reserves (dotted blue), total worst-case sink variability (dashed black), total actual sink variability (solid red). }
\label{fig:clustering_profiles}
\end{figure}

\begin{figure}[h]
\centering
\begin{tabular}{cc}
	\includegraphics[width=0.45\linewidth]{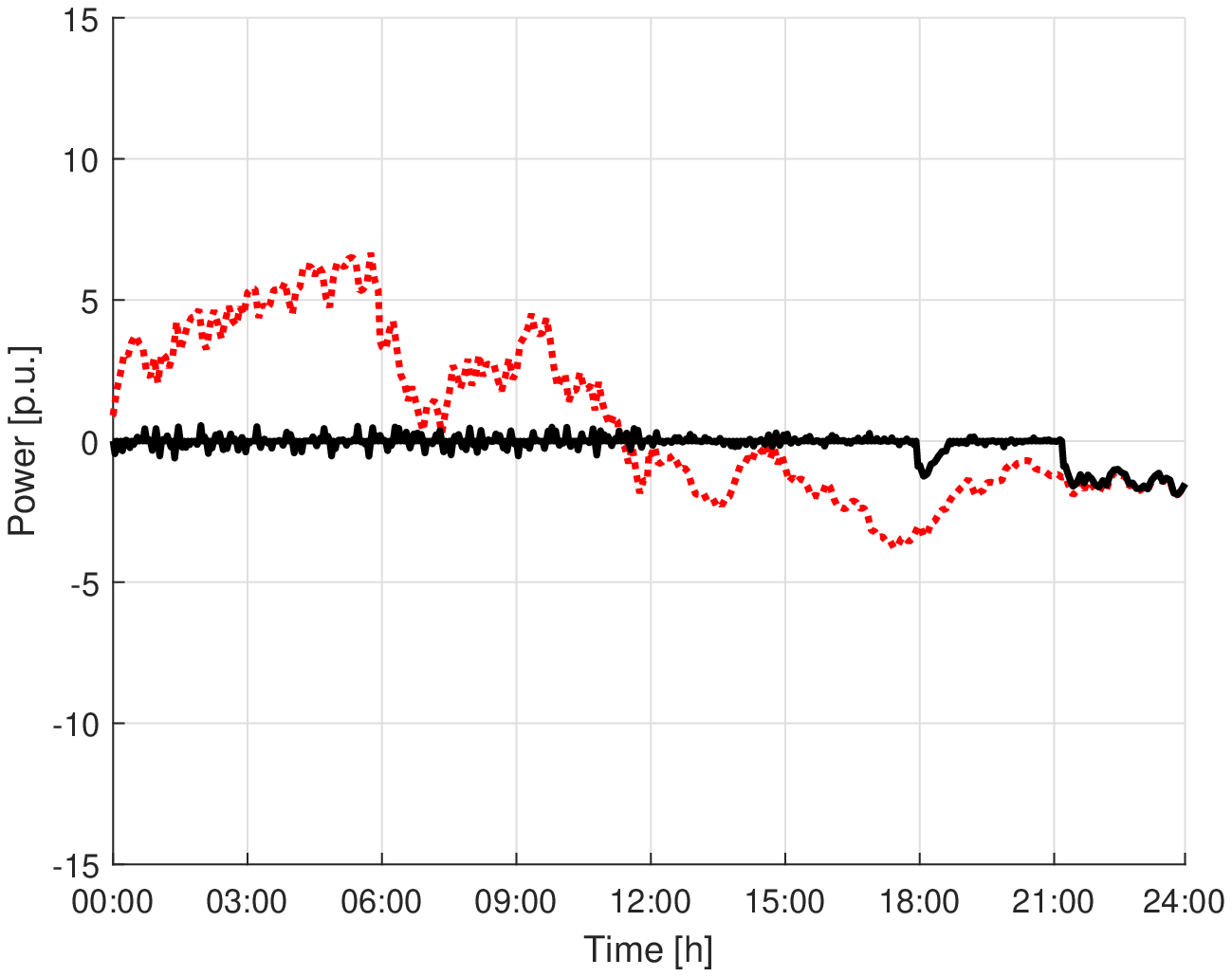} & \includegraphics[width=0.45\linewidth]{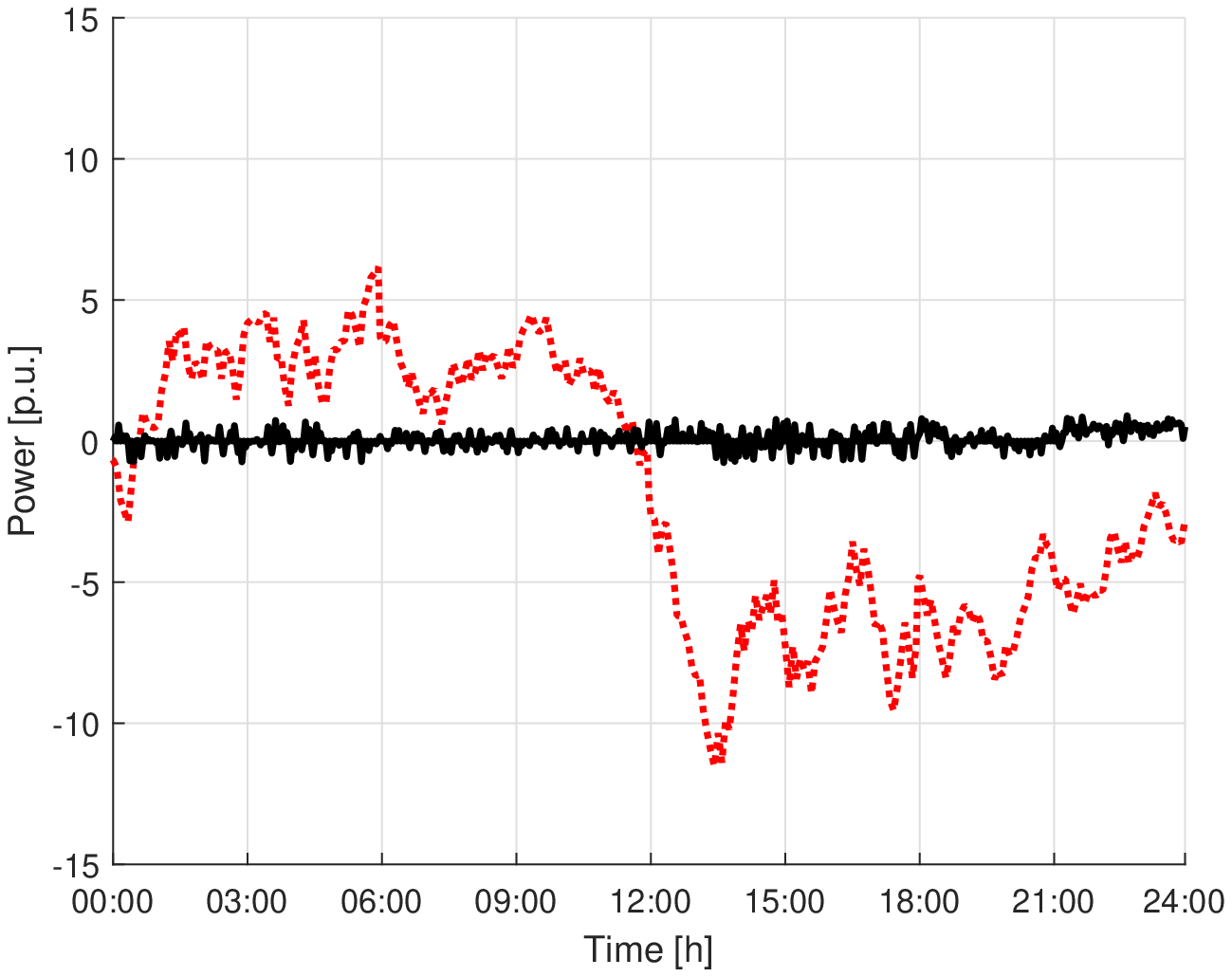}
	\\  (a) & (b) \\
	\includegraphics[width=0.45\linewidth]{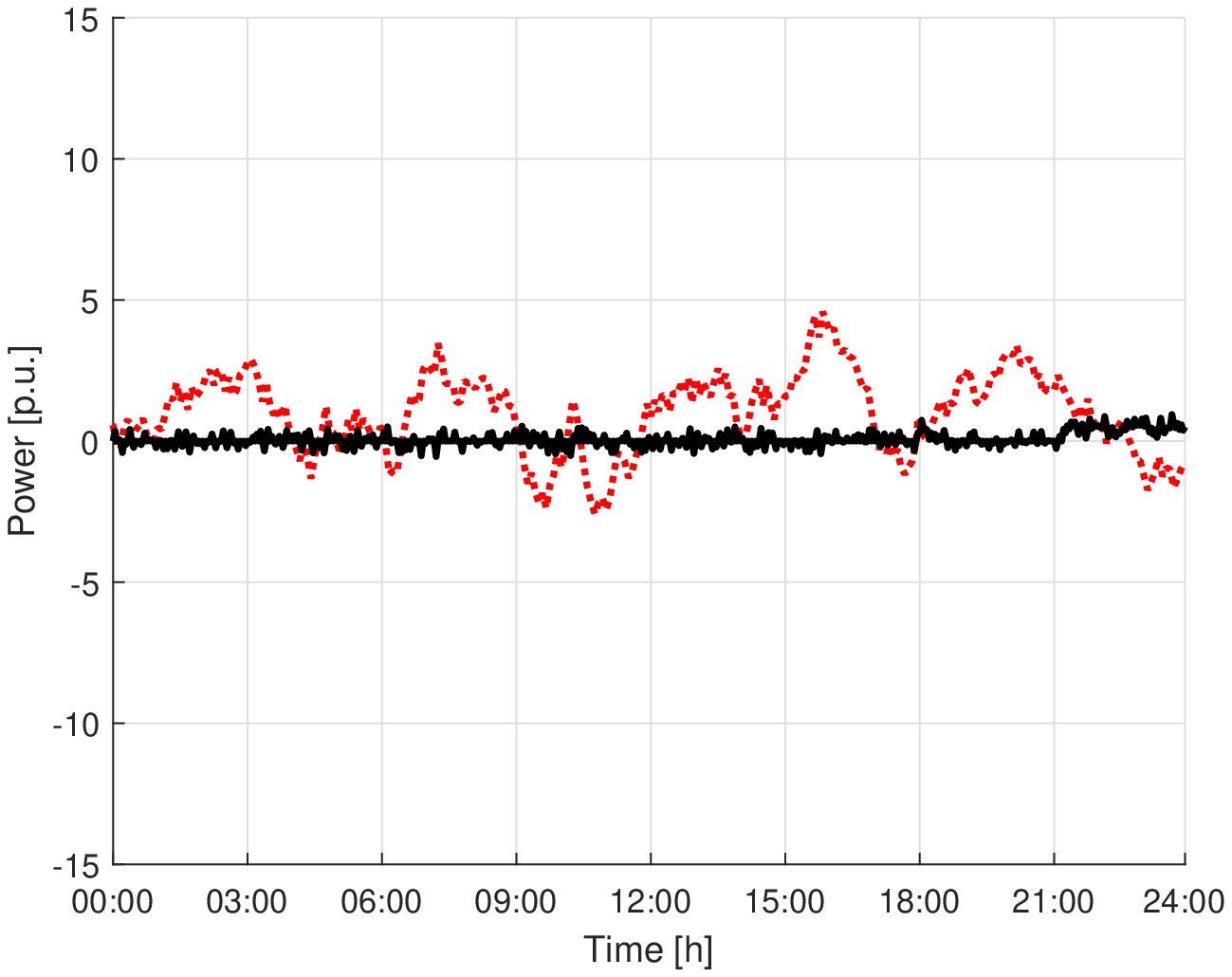} & \includegraphics[width=0.45\linewidth]{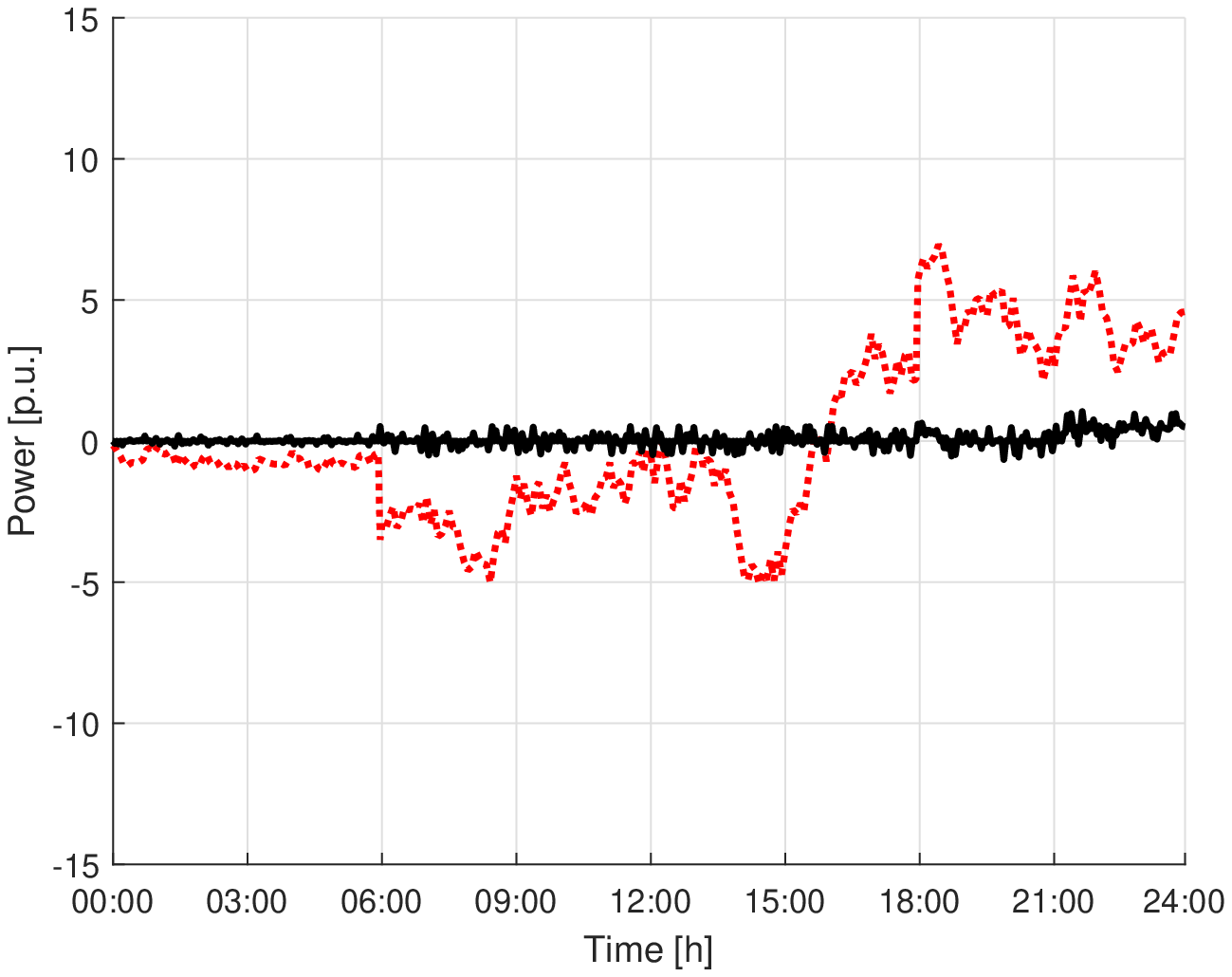}
	\\	(c) & (d) 
\end{tabular}
\caption{(a) Cluster 1, (b) Cluster 2, (c) Cluster 3 and (d) Cluster 4: output power deviation with respect to the pre-scheduled/forecasted one, in case the two-layer control is applied (solid black) and in case it is not (dotted red), i.e. where sources are controlled just to track the nominal profiles.}
\label{fig:variability}
\end{figure}

\begin{figure}[h]
\centering
\begin{tabular}{cc}
	\includegraphics[width=0.45\linewidth]{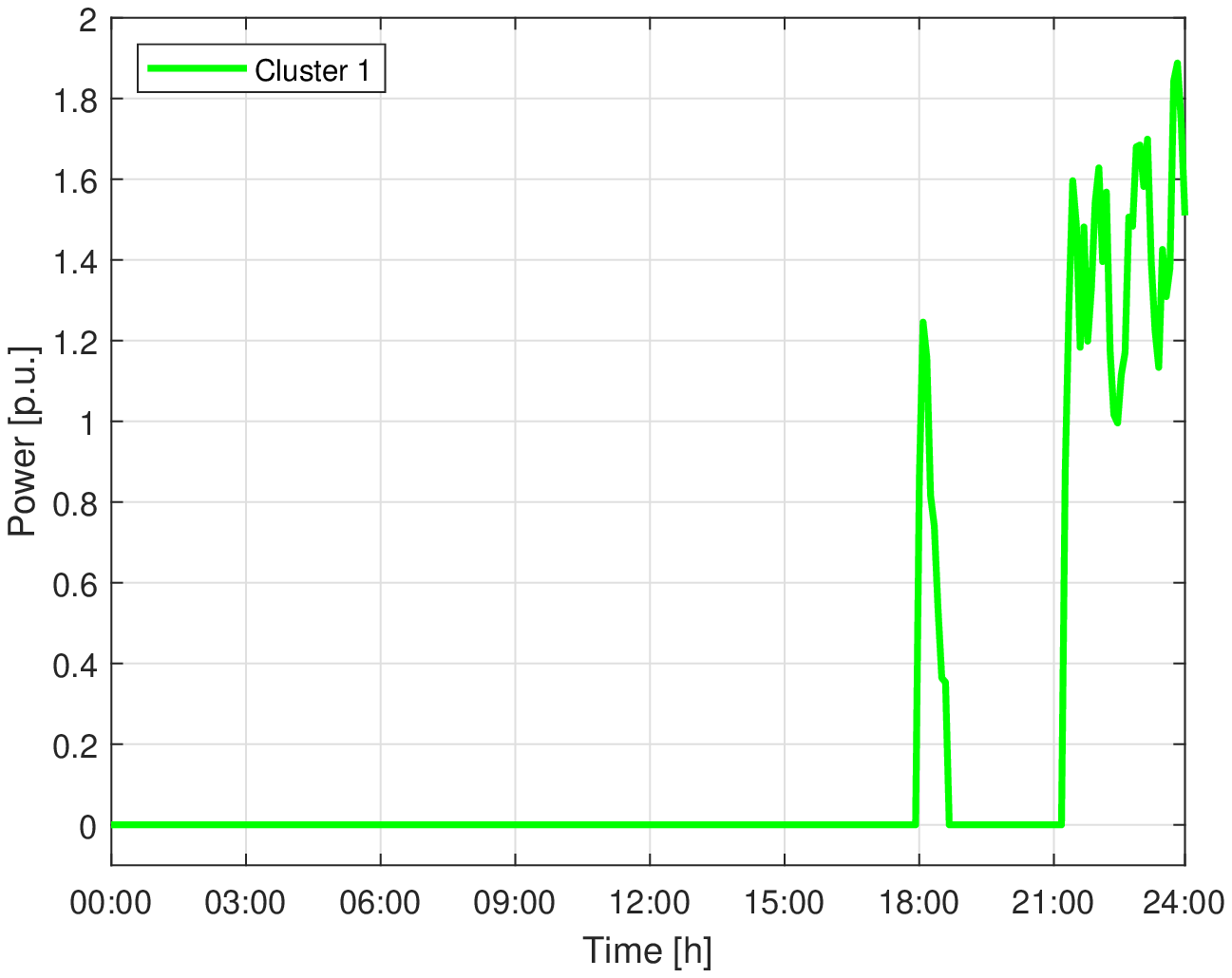} & \includegraphics[width=0.45\linewidth]{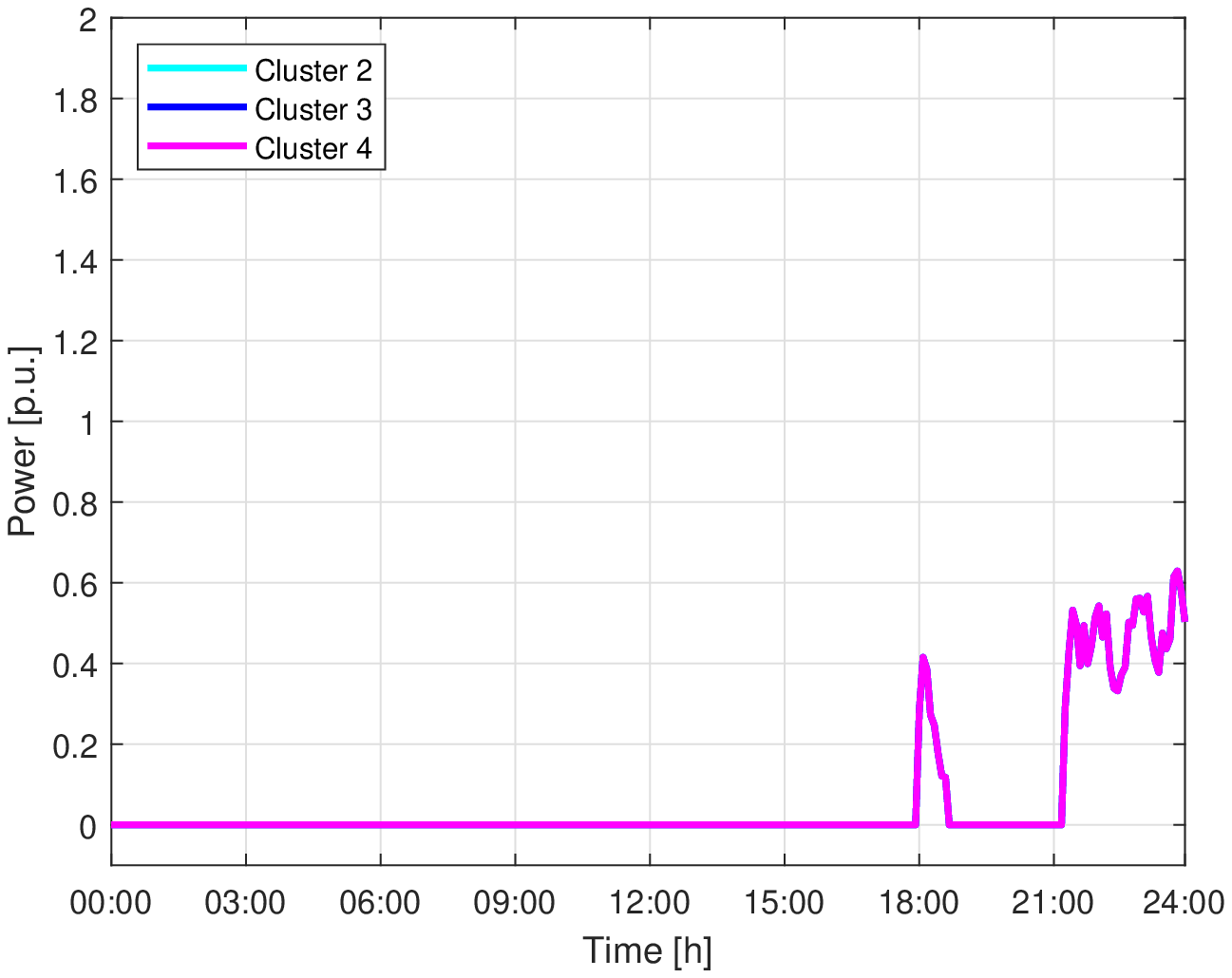}\\
	(a) & (b)
\end{tabular}
\caption{(a) Power requests transmitted by cluster MPC regulators; (b) Clusters' output power variation committed by supervisory layer.}
\label{fig:requests}
\end{figure}

\begin{figure}[h]
\centering
\includegraphics[width=0.45\linewidth]{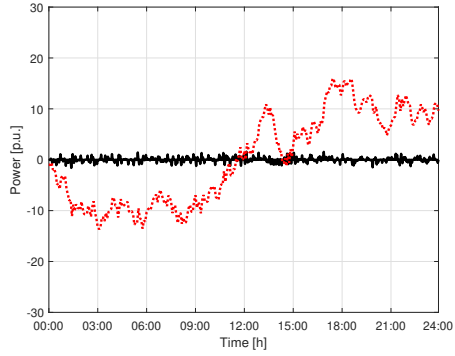}
\caption{Network total output power deviation if two-layer control is applied (solid black) and if not (dotted red).}
\label{fig:ieee118_variability}
\end{figure}

\begin{figure}[h]
\centering
\begin{tabular}{cc}
	\includegraphics[width=0.45\linewidth]{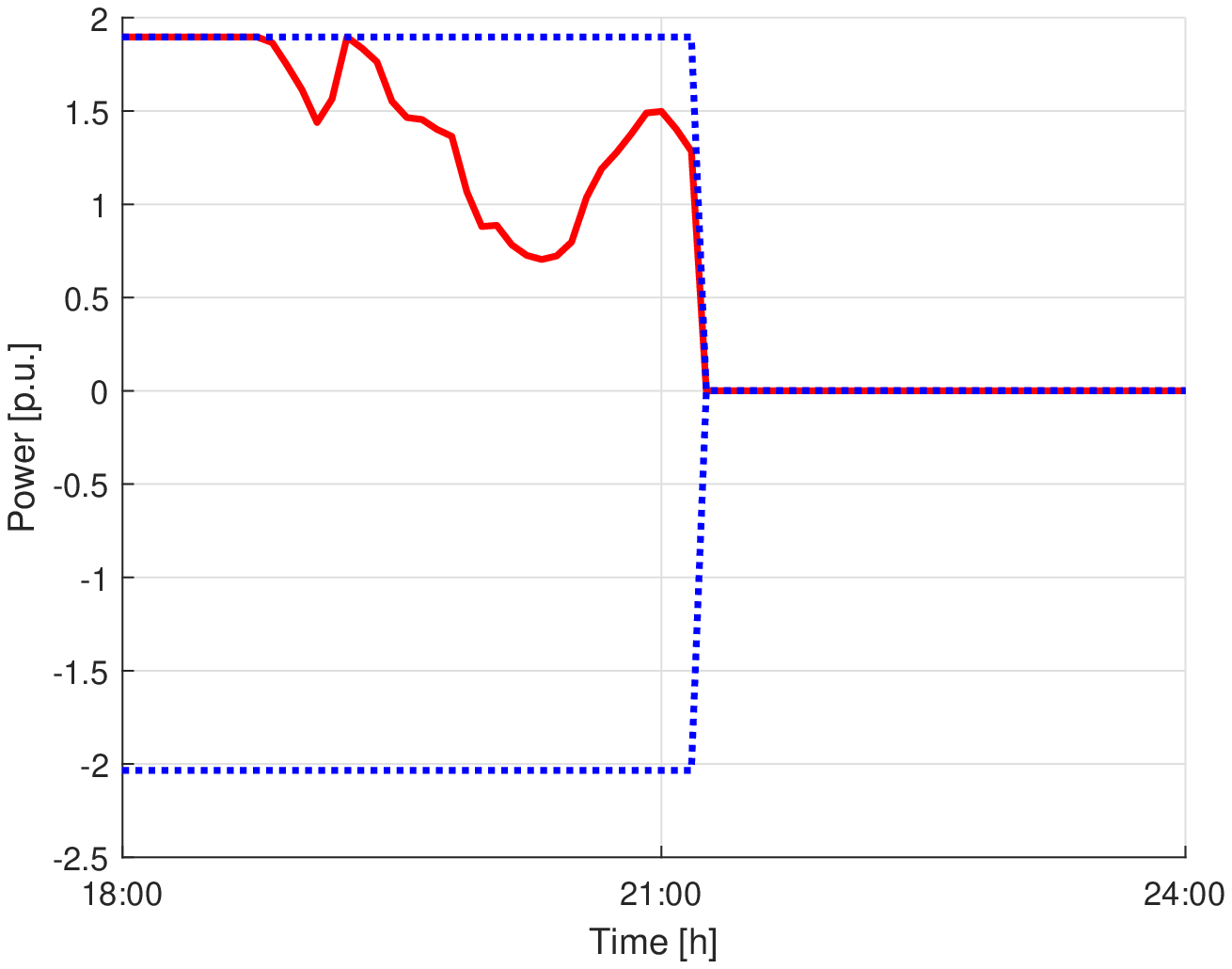} & \includegraphics[width=0.45\linewidth]{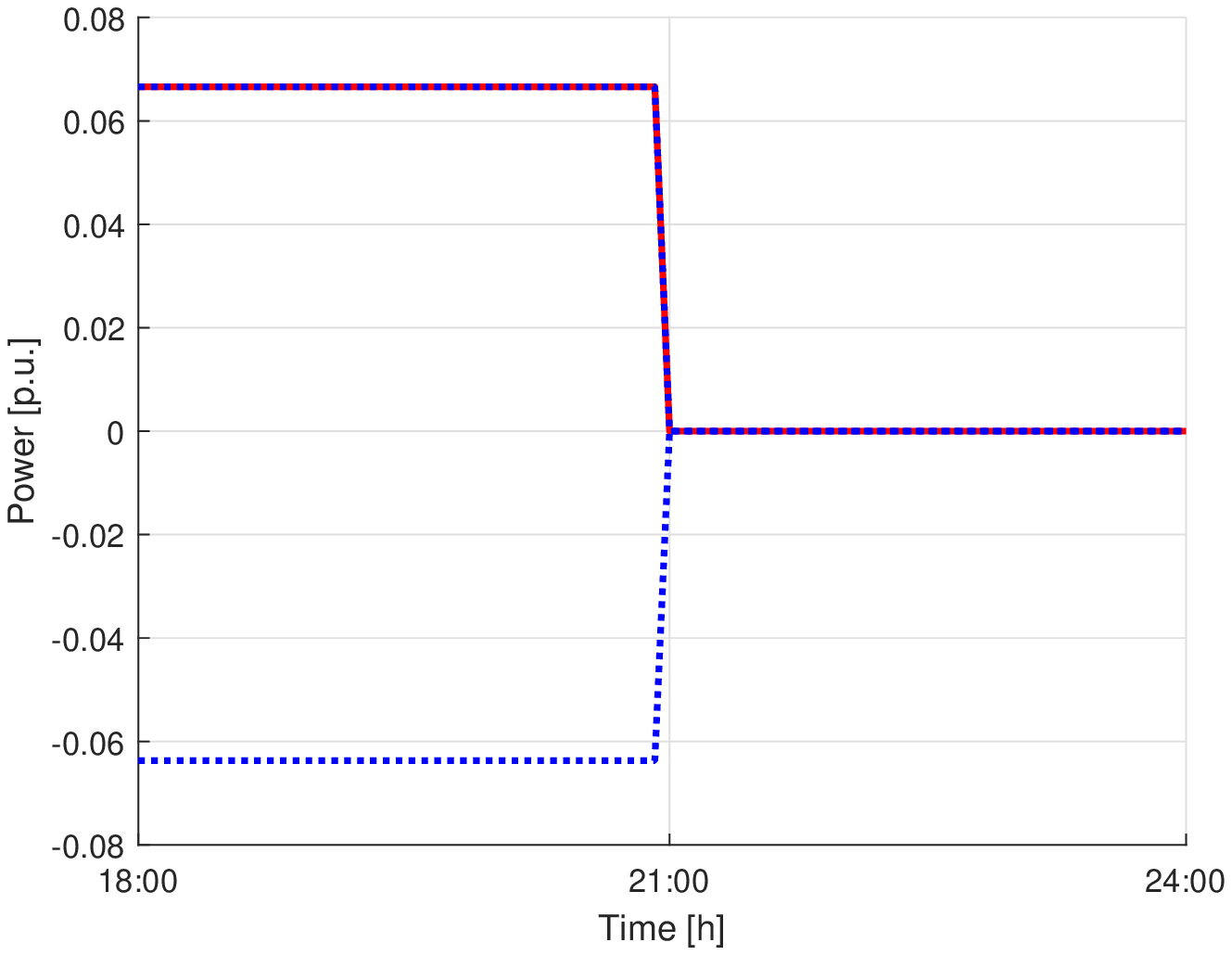}
	\\  (a) & (b) \\
	\includegraphics[width=0.45\linewidth]{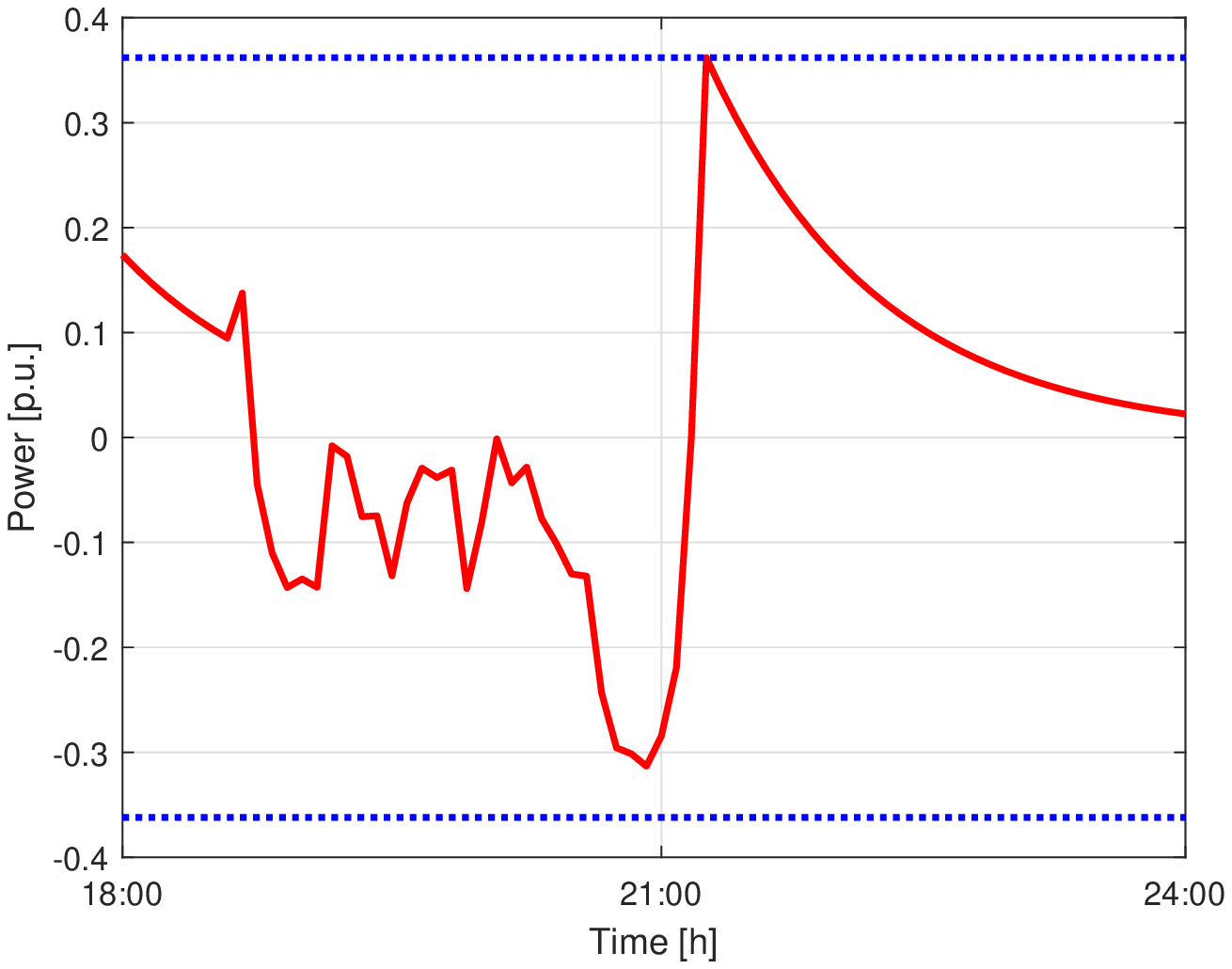} & \includegraphics[width=0.45\linewidth]{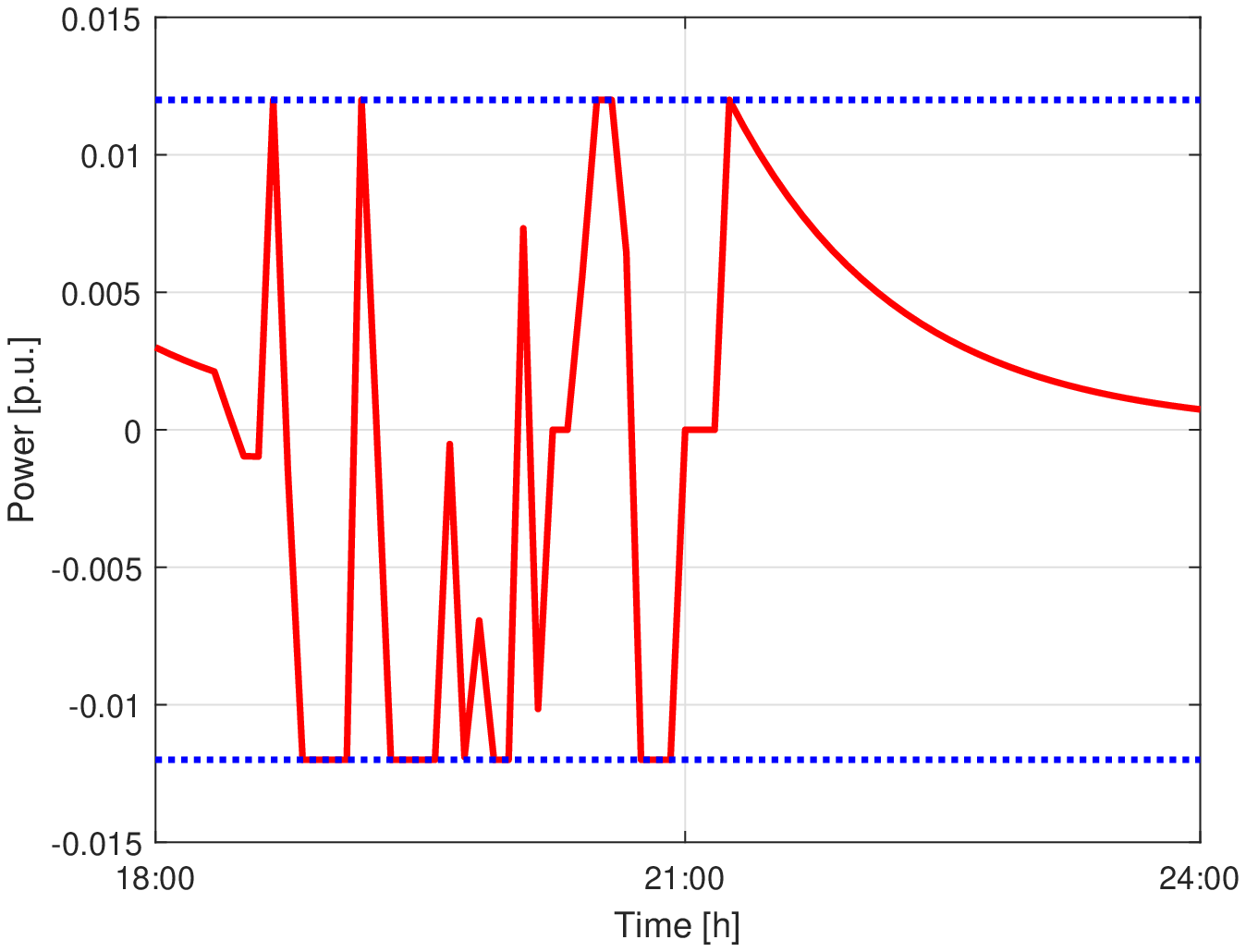}
	\\	(c) & (d) 
\end{tabular}
\caption{Local MPC regulation in Cluster 1 of (a) Source at node 3, (b) Source at node 5, (c) Battery at node 3 (d) Battery at node 5: output power variation (solid red), upward/downward reserves (dotted blue).}
\label{fig:gen_units}
\end{figure}

The control and clustering algorithms are defined considering $\tau=5 $ min and an overall time period of one day, implying that $N_t=24\text{h}/\tau=288$.
The described clustering procedure is carried out every 6 hours, meaning that $N_c=6\text{h}/\tau=72$ and $\gamma= N_t/N_c=4$.  Therefore, for the whole-daily management of the electrical network, the actual clusters will change their shape four times based on the availability of sources and on the variability of sinks. After having solved \eqref{eq:robust_up} and \eqref{eq:robust_dwn} to compute the optimal transactions, and having updated the graph weights as in \eqref{eq:reg_resource_from_proj}, the METIS partitioning tool is applied with $M=4$. Figure \ref{fig:clustering} shows the four clusters for each clustering period. Figure \ref{fig:clustering_profiles} reports the total sources' capability limits, the total worst-case and the real sinks' demand variability for each single cluster over the overall time period. As evident, the proposed clustering algorithm tries to partition the network such that  sources are sufficient to balance the sinks' demand variability in each cluster. This, however, is not the case for cluster 1 during the last clustering period, where the demand variability exceeds the capability limits of local sources.
\\
After having defined the network partitions for the different clustering periods, the proposed two-layer control architecture is applied. The local MPC layer is implemented  with a prediction horizon $N_p=15$. The distributed supervisory layer is implemented using an undirected, connected and complete communication graph  $\mathcal{G}_c=\{\mathcal{V}_c,\,\mathcal{E}_c\}$, meaning that each cluster can communicate with all the others.
\\
The variation of the clusters' output powers with respect to the pre-scheduled/nominal power flows are depicted in Figure \ref{fig:variability}, comparing the case where the proposed control architecture is implemented and the case it is not, i.e. when sources are not manipulated to compensate the power variability but to track the pre-scheduled nominal profiles. One can notice that, thanks to the efficient clustering procedure, sources are able to balance the sinks' demand variability in each cluster for most of the day, since the cluster's output power variation is mostly zero. However, after 18:00, the output power of cluster 1 shows a deviation related to the fact that local sources are not sufficient to balance the local variability (see Figure \ref{fig:variability}(a)). Indeed, the local MPC regulator of cluster 1 sends supporting requests to the supervisory layer between 18:00 and 24:00, as shown in Figure \ref{fig:requests}(a). These requests activate the distributed supervisory layer, which commits the needed power to the remaining clusters, i.e. cluster 2, 3 and 4, as evident from  Figure \ref{fig:requests}(b). It can be noticed from Figures \ref{fig:variability}(b)-(d) that the output power of clusters 2-3-4 increases after 18:00, so as to compensate cluster 1 negative deviation. As a result, the overall network variability is balanced at each time instant of the day, as shown in Figure \ref{fig:ieee118_variability}, meaning that the nominal power exchange with the main utility is maintained.  This balancing action is achieved by an accurate control of the local sources in each cluster, varying the output power compatibly with the offered reserves. Figure \ref{fig:gen_units} reports the local MPC action at cluster 1 for the local sources during the last clustering period.
\\
Finally, it is highlighted that the supervisory layer has been activated 41 times between 18:00 and 24:00, and the DC-ADMM required an average number of 33 iterations (min. 32, max. 35) to converge to the optimal solution. On the computer\footnote{A $5^{\text{th}}$ generation Lenovo ThinkPad X1 Carbon, Intel Core i7-7500U CPU @ 2.70 GHz, 16 GB RAM} it was run, the average execution time was 18.6 seconds (min. 17.7 s, max. 21.6 s). Neglecting the communication overhead, this amounts to roughly 4.7 seconds if the execution were performed in parallel by the four clusters, largely lower than the sampling time $\tau_s= 5$ min.

\vspace*{-20mm}

\clearpage
\section{Conclusions}
In this paper, an approach for the control of large-scale networked electricity systems has been presented. The main idea is to reduce the complexity of the problem by first partitioning the network into clusters. Then, for each cluster a local MPC regulator is designed with the aim of compensating local imbalances. If the generation resources in a cluster are not sufficient, a supervisory layer manages the interactions among clusters to guarantee the overall balancing.\\
Many improvements and variants of this method can be foreseen. Among them, its extension to water and heat distribution networks, as well as the inclusion of probabilistic forecasts for non dispatchable generators (renewables) and loads.
\vspace*{-2mm}

\section{Acknowledgements}
This work has received support from the Swiss National Science Foundation under the COFLEX project, Grant number 200021-169906. Riccardo Scattolini acknowledges the partial financial support by the Italian Ministry
for Research in the framework of the 2017 Program for Research Projects of National Interest (PRIN), Grant number 2017YKXYXJ. 
{The work of Alessio La Bella has been financed by the Research Fund for the Italian Electrical System in compliance with the Decree of Minister of Economic Development April 16, 2018.}
\vspace*{-2mm}

\bibliographystyle{IEEEtran}
\bibliography{IEEEabrv,articleref,thesis}

\small
\thispagestyle{empty}

\begin{wrapfigure}{l}{27mm} 
	\includegraphics[width=1in,height=1.25in,clip]{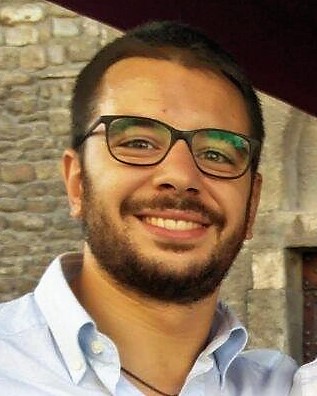}
\end{wrapfigure}
\par
\textbf{Alessio La Bella} received the Master of Science in Automation and Control Engineering from Politecnico di Milano in 2015. He received the Alta Scuola Politecnica Diploma, together with the Master of Science in Mechatronics Engineering from Politecnico di Torino in 2016. From January to October 2016, he worked at the research center RSE SpA - Ricerca sul Sistem Energetico, designing electrical simulators for assessing the integration of renewable sources in small islands. From November 2016 to October 2019, he was enrolled as a Ph.D. candidate in Information Technology at Politecnico di Milano, receiving the Ph.D. degree in February 2020. From September to December 2018, he was a Visiting Researcher at the Automatic Control Lab of the \'Ecole Polytechnique F\'ed\'erale de Lausanne, Switzerland. Since November 2019, he has been a post-doctoral researcher at the Department of Electronics, Information and Bioengineering of Politecnico di Milano. His main research interests concern the design of optimization-based control algorithms and machine learning techniques for the efficient management of energy and electrical systems.
\par
\vspace*{7mm}
\begin{wrapfigure}{l}{27mm}
	\vspace*{-5mm} 
	\includegraphics[width=1in,height=1.25in,clip]{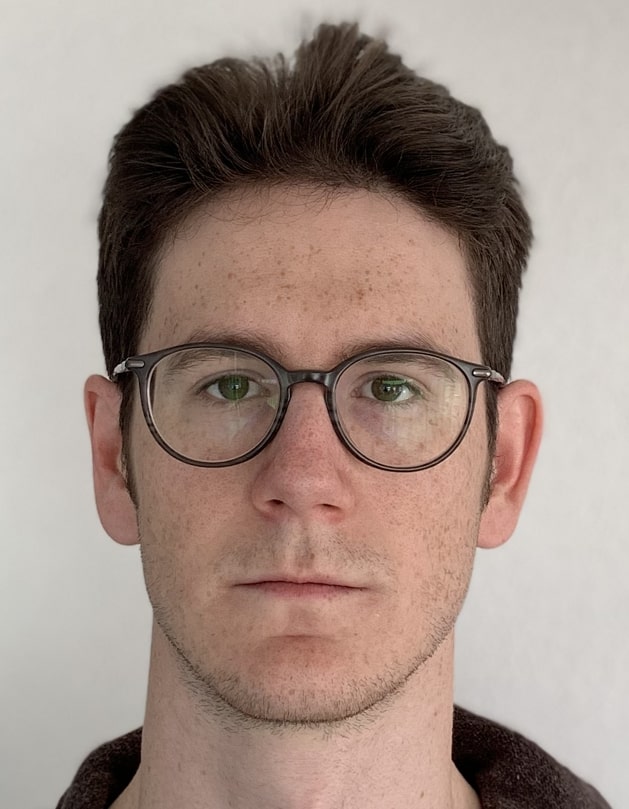}
\end{wrapfigure}
\par
\textbf{Pascal Klaus} received his Master of Science in Electrical and Electronic Engineering from the \'Ecole Polytechnique F\'ed\'erale de Lausanne in 2019. From January to July 2019, he spent a research period  at the Department of Electronics, Information and Bioengineering of Politecnico di Milano. He currently works as an MEP Engineer with Lombardi consutling engineers in Switzerland. His main interests include generation and distribution of electrical energy and process automation.\\ \\ \par \vspace*{5mm}  
\begin{wrapfigure}{l}{27mm}
	\vspace*{-5mm} 
	\includegraphics[width=1in,height=1.25in,clip]{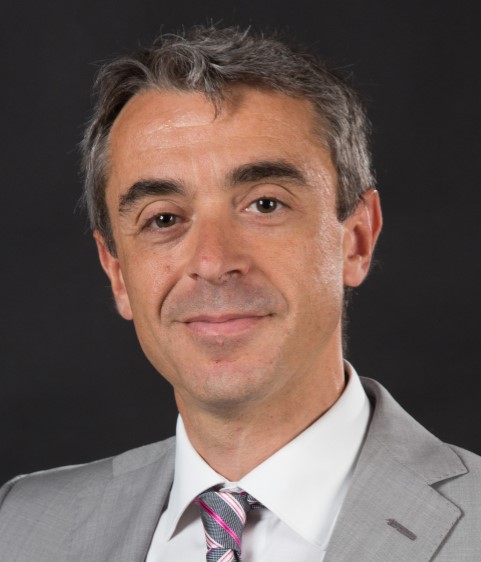}
\end{wrapfigure}
\par \textbf{Giancarlo Ferrari-Trecate} received the Ph.D. degree in Electronic and Computer Engineering from the Universita' degli  Studi di Pavia in 1999. Since September 2016 he is Professor at EPFL, Lausanne, Switzerland. In spring 1998, he was a Visiting Researcher at the Neural Computing Research Group, University of Birmingham, UK. In fall 1998, he joined as a Postdoctoral Fellow the Automatic Control Laboratory, ETH, Zurich, Switzerland. He was appointed Oberassistent at ETH, in 2000. In 2002, he joined INRIA, Rocquencourt, France, as a Research Fellow. From March to October 2005, he was researcher at the Politecnico di Milano, Italy. From 2005 to August 2016, he was Associate Professor at the Dipartimento di Ingegneria Industriale e dell'Informazione of the Universita' degli Studi di Pavia. His research interests include scalable control, microgrids, networked control systems, hybrid systems and machine learning.
Giancarlo Ferrari Trecate was the recipient of the Researcher Mobility Grant from the Italian Ministry of Education, University and Research in 2005. He is currently a member of the IFAC Technical Committees on Control Design and Optimal Control, and the Technical Committee on Systems Biology of the IEEE SMC society. He has been Editor at large for the 2019 ACC and has been serving on the editorial board of Automatica for three terms and of Nonlinear Analysis: Hybrid Systems. \par 
\vspace*{8mm}
\begin{wrapfigure}{l}{27mm}
		\vspace*{-5mm} 
	\includegraphics[width=1in,height=1.25in,clip]{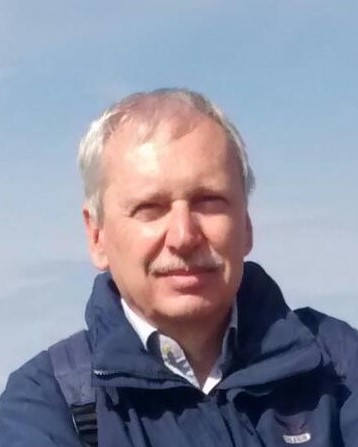}
\end{wrapfigure}
\par
\textbf{Riccardo Scattolini} is Full Professor of Automatic Control at the Politecnico di Milano, Italy. He was awarded Heaviside Premium of the Institution of Electrical Engineers. United Kingdom and was Associate Editor of the IFAC journal Automatica. His main research interests include modeling, identification, simulation, diagnosis, and control of industrial plants and electrical systems, with emphasis on the theory and applications of Model Predictive Control and fault detection methods to large-scale and networked systems.
\par
\end{document}